\documentclass[useAMS,usenatbib]{mn2e}
\usepackage{amssymb}
\usepackage{graphicx}
\usepackage{xspace}

\usepackage{amssymb,latexsym,graphicx,natbib,eufrak,times,amsmath}

\usepackage{ifthen}
\def\draftversion{1} % set this to 1 to display the content of the draft commands, or to 0 to hide them.

\makeatletter
\newcommand\mytoc{%
    \@starttoc{toc}%
}
\makeatother

\ifthenelse{\equal{\draftversion}{1}}{
	\usepackage{xcolor}
	\newcommand{\tmp}{}
	\newenvironment{envcomm}[1]{\renewcommand{\tmp}{#1}\begin{color}{blue}\begin{center}\hrule\vspace{0.5mm}\tmp's COMMENTS\end{center}}{\begin{center}END OF \tmp's COMMENTS\vspace{0.5mm}\hrule\end{center}\end{color}}
	\newenvironment{draft}{\begin{color}[rgb]{0,0.4,0}\begin{center}\hrule\vspace{0.5mm}DRAFT\end{center}}{\begin{center}END OF DRAFT\vspace{0.5mm}\hrule\end{center}\end{color}}
	\newcommand{\comcomm}[2]{\begin{color}{blue}\ $\bullet$ \textbf{#1:} #2 $\bullet$\ \end{color}}
	\newcommand{\revend}[1]{\par\begin{color}[rgb]{0,0.4,0}\begin{center}\hrule\vspace{0.5mm}END OF #1's REVISIONS\vspace{0.5mm}\hrule\end{center}\end{color}\par}
	\newcommand{\todo}[1]{\begin{color}{red}\ $\bullet$ \textbf{To do: }#1 $\bullet$\ \end{color}}
	
	\newcommand{\del}[1]{\begin{color}[rgb]{0,0.5,0.0}\ $\bullet$ \textbf{Deleted: }#1 $\bullet$\ \end{color}}
	\newcommand{\sk}[1]{\begin{color}[rgb]{0.6,0,0.6}#1\end{color}}
	\newcommand{\toc}{\par\begin{color}[rgb]{0.6,0,0.6}\begin{center}\hrule\vspace{0.5mm}\begingroup\small\let\cleardoublepage\relax\let\clearpage\relax\mytoc\endgroup\vspace{0.5mm}\hrule\end{center}\end{color}\par}
	}{
	\newsavebox{\trashcan}
	\newenvironment{envcomm}[1]{\begin{lrbox}{\trashcan}\begin{minipage}{\columnwidth}}{\end{minipage}\end{lrbox}}
	
	\newcommand{\comcomm}[2]{}
	\newcommand{\revend}[1]{}
	\newcommand{\todo}[1]{}
	
	\newcommand{\del}[1]{}
	\newcommand{\sk}[1]{}
	\newcommand{\toc}{}
	}
\long\def\symbolfootnote[#1]#2{\begingroup%
\def\thefootnote{\fnsymbol{footnote}}\footnote[#1]{#2}\endgroup} 

\newcommand{\aj}{AJ}% Astronomical Journal
\newcommand{\araa}{ARA\&A}% Annual Review of Astron and Astrophys
\newcommand{\apj}{ApJ}% Astrophysical Journal
\newcommand{\apjl}{ApJ}% Astrophysical Journal, Letters
\newcommand{\aap}{A\&A}% Astronomy and Astrophysics
\newcommand{\mnras}{MNRAS}% Monthly Notices of the RAS
\newcommand{\nat}{Nature}% Nature
\newcommand{\physrep}{Phys.~Rep.}% Physics Reports
\newcommand{\mh}{\ensuremath{\textrm{\,--\,}}}
\newcommand{\bb}[1]{\ifmmode \mbox{\boldmath $ #1$} \else \mbox{\boldmath $#1$} \fi}

\newcommand{\dd}{\ensuremath{\,\mathrm{d}}}

\newcommand{\U}[1]{\ensuremath{\mathrm{~#1}}}
\newcommand{\yr}{\U{yr}}
\newcommand{\Myr}{\U{Myr}}
\newcommand{\Gyr}{\U{Gyr}}
\newcommand{\pc}{\U{pc}}
\newcommand{\kpc}{\U{kpc}}
\newcommand{\Mpc}{\U{Mpc}}
\newcommand{\lsun}{\U{L}_{\odot}}
\newcommand{\Lsun}{\lsun}
\newcommand{\msun}{\U{M}_{\odot}}
\newcommand{\Msun}{\msun}
\newcommand{\Msunyr}{\Msun\yr^{-1}}
\newcommand{\cc}{\U{cm^{-3}}}

\newcommand{\kms}{\U{km\ s^{-1}}}

\newcommand{\hi}{H{\sc i} }
\newcommand{\hii}{H{\sc ii} }

\newcommand{\ramses}{\texttt{RAMSES}\xspace}
\newcommand{\por}{\texttt{POR}\xspace}
\newcommand{\nemo}{\texttt{NEMO}\xspace}

\newcommand{\fig}[2][]{Figure#1~\ref{fig:#2}}
\newcommand{\tab}[2][]{Table#1~\ref{tab:#2}}
\newcommand{\sect}[2][]{Section#1~\ref{sec:#2}}
\newcommand{\app}[2][]{Appendix#1~\ref{sec:#2}}

\renewcommand{\fig}[2][]{Fig#1.~\ref{fig:#2}}

\newcommand{\newta}{N$_\textrm{A}$\xspace}
\newcommand{\monda}{M$_\textrm{A}$\xspace}
\newcommand{\newts}{N$_\textrm{S}$\xspace}
\newcommand{\monds}{M$_\textrm{S}$\xspace}

\title[Star formation with modified gravity]{Star formation triggered by galaxy interactions in modified gravity}

\author[Renaud, Famaey \& Kroupa] {Florent~Renaud$^1$\thanks{f.renaud@surrey.ac.uk}, Benoit~Famaey$^2$ \& Pavel~Kroupa$^{3,4}$\\
$^1$Department of Physics, University of Surrey, Guildford, GU2 7XH, UK\\
$^2$Observatoire Astronomique de Strasbourg, Universit\'e de Strasbourg, CNRS UMR 7550, 11 rue de l'Universit\'é, F-67000 Strasbourg, France\\
$^3$Helmholtz-Institut f\"ur Strahlen- und Kernphysik, Nussallee 14–16, D-53115 Bonn, Germany\\
$^4$Charles University in Prague, Faculty of Mathematics and Physics, Astronomical Institute, V Hole\v{s}ovi\v{c}k\'ach 2, CZ-180 00 Praha 8, Czech Republic
}

\date{Accepted 2016 September 12. Received 2016 August 24; in original form 2016 July 24}

\begin{document}
\maketitle

%%%%%%%%%%%%%%%%%%%%%%%%%%%%%%%%%%%%%%%%%%%%%%%%%%%%%%%%%%%%%%%%%%%%%%%%%%%%%%%%

\begin{abstract}
Together with interstellar turbulence, gravitation is one key player in star formation. It acts both at galactic scales in the assembly of gas into dense clouds, and inside those structures for their collapse and the formation of pre-stellar cores. To understand to what extent the large scale dynamics govern the star formation activity of galaxies, we present hydrodynamical simulations in which we generalise the behaviour of gravity to make it differ from Newtonian dynamics in the low acceleration regime. We focus on the extreme cases of interacting galaxies, and compare the evolution of galaxy pairs in the dark matter paradigm to that in the Milgromian Dynamics (MOND) framework. Following up on the seminal work by Tiret \& Combes, this paper documents the first simulations of galaxy encounters in MOND with a detailed Eulerian hydrodynamical treatment of baryonic physics, including star formation and stellar feedback. We show that similar morphologies of the interacting systems can be produced by both the dark matter and MOND formalisms, but require a much slower orbital velocity in the MOND case. Furthermore, we find that the star formation activity and history are significantly more extended in space and time in MOND interactions, in particular in the tidal debris. Such differences could be used as observational diagnostics and make interacting galaxies prime objects in the study of the nature of gravitation at galactic scales.
\end{abstract}

%%%%%%%%%%%%%%%%%%%%%%%%%%%%%%%%%%%%%%%%%%%%%%%%%%%%%%%%%%%%%%%%%%%%%%%%%%%%%%%%
\section{Introduction}

Interactions mark milestones in the evolution of galaxies by modifying their mass, stellar, gaseous and chemical contents, morphology, kinematics and dynamical properties \citep[see e.g.][among many others]{Arp1966, Sanders1996, Springel1999, Struck1999, Saintonge2012, Duc2013}. These events are often (but not always) associated with burst(s) of star formation such that, in the local Universe, all the most luminous galaxies (e.g. $> 10^{12} \Lsun$ for the ultra luminous infrared galaxies, ULIRGs, \citealt{Houck1985, Kennicutt1998b}) yield the signatures of major interactions \citep{Armus1987, Ellison2013}. Numbers of studies in all wavebands have characterised the properties of interacting systems, in particular their star formation activity, with the aim of pinning down the underlying physical processes \citep[e.g.][]{Schombert1990, Hibbard1995, Bournaud2004, Chien2007, Smith2010, Boquien2011, Saintonge2012, Scudder2012}.

In these fast evolving objects with complex geometries, numerical simulations have long been considered as a fundamental complement to observations. Starting with \citet{Toomre1972}, all works point out the paramount role of gravitation on affecting both the galactic scale structures \citep[e.g.][]{Barnes1992, Quinn1993, Dubinski1996, Mihos1998, DiMatteo2007, Hopkins2009, Renaud2009, Moreno2013, Privon2013} and the internal, small scale, physics of interacting galaxies \citep[e.g.][]{Barnes1991, Teyssier2010, Chien2010, Hopkins2013, Renaud2014b}. The scale-free aspect of gravitation makes it indeed a key process at galactic scale in the shaping of galaxies and their structures (spiral, bars, large scale flows), at sub-galactic scales in the assembly of molecular clouds \citep{MacLow2004, Hennebelle2008} and down to the scale of pre-stellar cores in star formation \citep{Bate2005, Bonnell2013}. Galaxy interactions are thus the perfect benchmark to understand the role of gravitation on star formation, at both galactic and sub-galactic scales.

The classical framework in which theoretical galactic studies are performed these days is the $\Lambda$-Cold Dark Matter ($\Lambda$CDM) paradigm. However, both the cosmological constant $\Lambda$ and the CDM part of the model could also be related to a modification of gravity. On galaxy scales, the model is indeed plagued by severe problems, the most famous ones being the cusp-core problem (\citealt{deBlok2010, Oman2015}, but see also \citealt{Read2016b}), the too-big-to fail problem \citep{BoylanKolchin2011, Papastergis2015, Pawlowski2015}, or the satellite planes problem \citep{Kroupa2005b, Metz2007, Metz2008, Pawlowski2012, Ibata2013, Ibata2014, Pawlowski2015}. There is also a more general problem linked to the finely tuned relation between the distribution of baryons and the gravitational field in galaxies, as encapsulated in various scaling relations involving a universal acceleration constant $a_0 \approx 10^{-10} \U{m\ s^{-2}}$, including the tight baryonic Tully-Fisher relation \citep{McGaugh2000, Lelli2016a, Papastergis2016}, the diversity of shapes of rotation curves at a given maximum velocity scale \citep{Oman2015}, or the relation between the stellar and dynamical surface densities in the central regions of galaxies \citep{Lelli2016b, Milgrom2016}, and many others \citep{Famaey2012}. All this points to things happening as if the effects usually attributed to CDM on galaxy scales were actually due to a modified force law. The {\it a priori} simplest explanation for this would be that gravity is indeed effectively different in the weak field regime and accounts for the effects usually attributed to CDM. This paradigm is known as Modified Newtonian Dynamics (MOND), or Milgromian Dynamics, suggested more than 30 years ago by \citet{Milgrom1983}. It predicted all the observed galaxy scaling relations well before they were precisely assessed by observations \citep{Famaey2012}. Nevertheless, this paradigm cannot be complete, as a full theory of gravitation, also valid on cosmological scales, has not yet been found. But while successful on galaxy scales, the MOND paradigm has still been far from being fully explored even on these scales where it is currently successful. Hence there is still potential for falsification of this paradigm in its {\it a priori} domain of validity. The main reason for this lack of exploration of all predictions of MOND on galaxy scales is its non-linear nature, and the previous lack of numerical codes devised to model galaxies in this framework.

After the pioneering work of \citet{Brada1999}, only a handful of codes have been designed \citep{Nipoti2007, Tiret2007, Llinares2008, Angus2012}, but all with their own caveats, notably regarding the treatment of hydrodynamics. The first treatment of gas in MOND simulations was proposed by \citet{Tiret2008a}, who used sticky particles, but a full hydrodynamical treatment of gas has become possible only recently thanks to the patches of the \ramses code developed by \citet{Lughausen2015} and \citet{Candlish2015}. This is particularly important in order to study star formation in general, and in galaxy interactions in particular. The role of gravitation on star formation is central at both galactic and sub-galactic scales, and modifying it must have potentially observational consequences. 

For instance, during interactions, tides can induce the formation of tails expanding up to $\sim 100 \kpc$ away from their progenitor galaxies. Some regions of these tails can become unstable, fragment and form stellar objects as massive as dwarf galaxies ($\sim 10^{8-9} \Msun$). However, self-gravity is usually too weak to assemble such tidal dwarf galaxies (TDGs), and an external contribution to the local potential well is required. With $\Lambda$-CDM, only the potential well of the galactic DM halo can have such catalyst effect and thus, it must be sufficiently extended to embed the TDGs along the long tidal tails \citep{Bournaud2003}. But since the tidal debris originates from the discs (and thus contains very little DM, if any) and because the surrounding DM halos are dynamically hot, the DM distribution does not follow the baryonic one, and their external effect on TDG seeds remains mild. In MOND however, the baryonic seeds of TDGs generate their own ``phantom dark matter'' and thus, an additional potential well. Therefore, instabilities leading to the formation of TDG seeds are strengthened by the local MOND potential which amplifies them and allow them to grow. As a consequence, the formation of TDGs is eased in the MOND framework, compared to the Newton case \citep{Tiret2008b,Combes2010}. Since it provides a test of the gravitation paradigm, the nature of observed TDG candidates is intensively debated, in particular in the context of the the Tully-Fisher relation \citep[see e.g.][]{Gentile2007, Lelli2015, Flores2016} and regarding the potential origin of satellite galaxies sitting on satellite planes in the context of MOND, which could actually be old TDGs instead of primordially formed dwarf galaxies \citep[e.g.][]{Kroupa2010, Kroupa2015}.

In this paper, we thus use the MOND framework to analyse the role of gravitation in enhancing star formation in interacting galaxies. We characterise the starburst activity associated to interactions in the context of MOND, and compare to that obtained when they are surrounded by halos of particle dark matter. We present our suite of simulations in Sect.~2 and the results in Sect.~3. Conclusions are drawn in Sect.~4.

%%%%%%%%%%%%%%%%%%%%%%%%%%%%%%%%%%%%%%%%%%%%%%%%%%%%%%%%%%%%%%%%%%%%%%%%%%%%%%%%
\section{Formalism and numerical method}

%%%%%%%%%%%
\subsection{Modified gravity}
\label{sec:mond}

In highly symmetrical situations (such as spherical symmetry) the net MOND gravitational acceleration $\bb{g}$ is connected to the Newtonian term $\bb{g}_\mathrm{N}$ through
\begin{equation}
\bb{g} = \nu\left(\frac{g_\mathrm{N}}{a_0}\right) \bb{g}_\mathrm{N},
\end{equation}
where $a_0$ is a constant acceleration, generally chosen to be $a_0 \approx 10^{-10} \U{m\ s^{-2}}$, and $\nu$ is an interpolation function such that
\begin{equation}
\left\{\begin{array}{ll}
\nu (x) \rightarrow 1 & \textrm{for}\quad x \gg 1 \quad\textrm{(Newtonian regime)}\\
\nu (x) \rightarrow x^{-1/2} &\textrm{for}\quad x \ll 1 \quad\textrm{(MOND regime)}\\
\end{array}
\right..
\end{equation}
In the weak field (or MOND) regime, one gets
\begin{equation}
\bb{g} \rightarrow \sqrt{ a_0 \bb{g}_\mathrm{N}},
\end{equation}
which successfully predicts the observed rotation curves of galaxies \citep[see][for a review]{Famaey2012}. This relation indeed predicts for instance that galaxies of the same baryonic mass would share the same asymptotic circular velocity in accordance with the baryonic Tully-Fisher relation, but also that the central slope of the rotation curve of disc galaxies scales as $(\dd V/ \dd R)_0 \propto \sqrt{\nu \rho_\mathrm{b}(0)}$, where $V$ is the circular velocity, $R$ is the galactocentric distance, $\rho_\mathrm{b}(0)$ is the central baryonic density (which is typically the central baryonic surface density over twice the scale-height) and $\nu$ is taken close to the centre (at the first measured point of the rotation curve). Since $\nu$ depends on $g_\mathrm{N}$, it also scales with the baryonic surface density, and a prediction of MOND is consequently that the central circular velocity slope depends mainly on the central baryonic surface density of the galaxy. Hence, MOND naturally predicts the diversity of rotation curve shapes at a given mass scale noted by \citet{Oman2015}, solely from the existence, at a given mass scale, of a diversity of central baryonic surface densities, and it also predicts a strong correlation of the latter with the circular velocity gradient $(\dd V/ \dd R)_0$, as observed by \citet{Lelli2013}.

Outside of spherical symmetry, the previous formula must nevertheless be modified. This is usually done by modifying the Poisson equation, following the least action principle for a modified Lagrangian of gravitation. One such flavour of MOND, called Quasi-Linear MOND \citep[QUMOND, ][]{Milgrom2010}, provides a formalism similar to the Newtonian case by introducing a ``phantom dark matter'' density which is fully determined by the baryon distribution.

Following \citet{Famaey2005}, for galaxies\footnote{See however \citet{Hees2016} for constraints in the Solar System.}, we can choose an interpolation function\footnote{This function is equivalent to the inverse interpolation function $\mu$ in Equation~2 of \citet{Candlish2015}.}
\begin{equation}
\label{eqn:nu}
\nu(x) = \frac{1}{2} + \frac{\sqrt{x^2+4x}}{2x}.
\end{equation}
Then, following \citet{Famaey2012}, we define the function\footnote{Our $\tilde{\nu}$ equals the function $\nu$ of \citet{Lughausen2015}.}
\begin{equation}
\tilde{\nu}(x) = \nu(x) - 1
\end{equation}
that allows us to write the modified Poisson equation
\begin{equation}
\nabla^2\phi(\bb{x}) = 4\pi G \rho_\mathrm{b}(\bb{x}) + \nabla.\left[\tilde{\nu}\left(\frac{|\nabla \phi_\mathrm{N} |}{a_0}\right) \nabla \phi_\mathrm{N}(\bb{x}) \right]
\end{equation}
where $\phi$ and $\phi_\mathrm{N}$ are the net and Newtonian potentials respectively, and $\rho_\mathrm{b}$ is the baryonic density, which fulfils the Newtonian Poisson equation $\nabla^2\phi_\mathrm{N}(\bb{x}) = 4\pi G \rho_\mathrm{b}(\bb{x})$. We then introduce the ``phantom dark matter'' density $\rho_\mathrm{ph}$
\begin{equation}
\label{eqn:poisson}
\nabla^2\phi(\bb{x}) = 4\pi G \left[\rho_\mathrm{b}(\bb{x}) + \rho_\mathrm{ph}(\bb{x}) \right],
\end{equation}
which is fully defined once the baryonic distribution is known, and can be seen as the (non-particle) MOND equivalent to the DM contribution in the classical case. In the rest of the paper, we adopt the QUMOND formalism and the standard value $a_0 = 1.12 \times 10^{-10} \U{m\ s^{-2}}$.

Since MOND has been intensively used in the last decades to mainly study the dynamics of the outer regions of galaxies and galaxy clusters, it is often wrongly seen as a modification of Newtonian dynamics at large scales ($\sim 100 \kpc \mh 1 \Mpc$). It is however important to keep in mind that MOND does not directly depend on spatial scale but on the local gravitational acceleration. Therefore, it plays an important role where Newtonian gravitation is weak (parametrised by $a_0$), which is not strictly equivalent as playing an important role in the outskirts of galaxies, or in galaxy clusters, like DM does.

%%%%%%%%%%%
\subsection{Numerical method}
\label{sec:method}

A handful of simulation codes have been developed within the MOND framework to address questions raised by Milgromian gravity \citep{Brada1999, Llinares2008, Nipoti2007, Tiret2008b, Candlish2015, Lughausen2015}. However, in the era of multi-scale and multi-physics numerical studies, it is necessary to include the generalisation of the gravitation law in a broader context, notably to correctly treat the hydrodynamics and the physics of star formation. That is why not only Poisson solvers but also more versatile simulation codes originally designed in the Newtonian framework have recently been modified to solve the generalised Poisson's equation of MOND. Most notably, \citet{Lughausen2015} proposed a patch named {\tt PHANTOM OF RAMSES} (\por) to the adaptive mesh refinement (AMR) \ramses code \citep{Teyssier2002}. No other functionalities or implementations of \ramses than the Poisson solver are affected by the use of \por. 

All the simulations presented here are performed using the adaptive mesh refinement \ramses code and the \por patch. Any other physical aspect is treated as in \citet{Renaud2015a}. In short, the galaxy systems (isolated or in pairs) are modelled without accretion of external gas nor DM. Heating comes from ultraviolet radiation of cosmic origin (tabulated at redshift zero as in the public version of \ramses) and stellar feedback (see below), and the atomic cooling used is tabulated at solar metallicity \citep{Courty2004}. Gas denser than a density threshold of $0.6 \cc$ is converted into star particles, at a fixed efficiency per free-fall time (4\%, see also \citealt{Renaud2013b}). The threshold combined with the efficiency corresponds to a SFR of the order of $1\Msunyr$ for isolated galaxies, i.e. typical of the main sequence galaxies in the local Universe \citep[e.g.][]{Daddi2010b}. Star formation is only active in the densest regions of the galaxies, i.e. the most refined volumes of the simulations. \app{refinement} shows that in these regions, small-scale dynamics are dominated by the baryonic component and are fairly independent of the gravitation paradigm adopted. The self-gravity of such systems, responsible for cloud collapse and fragmentation leading to star formation, lies in the non-modified regime, or strong field, of MOND. Therefore, we use the exact same sub-grid recipe for star formation for both the Newtonian and Milgromian runs. Differences in the star formation activities would thus originate from the formation of the star forming clouds themselves due to larger scale dynamics like large-scale gas flows, compression, shocks, shear, etc.

Stellar feedback includes photo-ionisation in \hii regions, radiation pressure \citep{Renaud2013b} and thermal type-II supernovae \citep{Dubois2008, Teyssier2013}. The simulated volume spans $(400 \kpc)^3$, with Dirichlet's boundary conditions, and with the highest resolution of the AMR grid being $6 \pc$. Additional tests show that our results are not affected by the choice of size for the simulation volume.

The simulations have been performed on the \emph{Curie} supercomputer hosted at the \emph{Tr\`es Grand Centre de Calcul} (TGCC).

%%%%%%%%%%%
\subsection{Simulation suite}
\label{sec:suite}

In the following, all galaxy models are the same (apart from the presence/absence of a DM halo), as described in \sect{ic}. The differences in the simulation parameters are solely based on the orbits. We have performed four simulations of interacting galaxies, as follows.
\begin{itemize}
\item Using the orbit of \citet{Renaud2015a} designed to reproduce the morphology and kinematics of the Antennae galaxies (NGC~4038/39), we run a simulation in the Newtonian framework using the MW-A model \citep{Kuijken1995} setup for both progenitors as described in \sect{ic}. Using this galaxy model introduces slight differences from \citet{Renaud2015a} in the final result, and the match to the Antennae system is, de facto, not as good as in previous works. This simulation is labelled \newta hereafter. The equivalent MOND simulation, labelled \monda, is setup using the exact same parameters: same orbit and same progenitors, only with the DM particles removed and replaced by the QUMOND gravitation, as detailed in Sections~\ref{sec:mond} and \ref{sec:ic}. The system obtained is thus \emph{not} tailored to reproduce any observation of the real Antennae. It however provides a comparison as direct as possible with the Newtonian case. For these simulations, the initial relative velocity of the progenitors is $\approx 160 \kms$.
\item We design another MOND simulation, named \monds (for ``slow''), providing a better fit to the Antennae although precisely matching all features of the real system is not our goal. Keeping the progenitors unchanged (for simplicity), the main objective of this run is to compensate for the differences in dynamical friction between Newton and MOND \citep[e.g.][]{Kroupa2015}, to reduce the orbital period and to adjust the length of the tidal tails (see below). All these targets have been reached by decreasing the initial orbital velocity of the progenitors by a factor of two (i.e. down to $\approx 80 \kms$)\footnote{Such velocities are un-representative of galaxy group and cluster environments but fully compatible with that of pairs \citep[see e.g.][]{Chou2012}.}. For the sake of completeness, we finally run the Newtontian equivalent (\newts) using the same initial parameters as in \monds.
\end{itemize}

\begin{table}
\caption{Initial orbital parameters}
\label{tab:ic_orbit}
\begin{tabular}{lcc}
\hline
Simulation & \newta and \monda & \newts and \monds \\
\hline
\multicolumn{3}{l}{Galaxy 1}\\
position [kpc] & (12.7,-30.3,46.7) & = \\
velocity [km/s] & (-26.9,23.3,-71.5) & $\times 0.5$\\
spin axis & (0.67,-0.71,0.20) & = \\
\hline
\multicolumn{3}{l}{Galaxy 2}\\
position [kpc] & (-12.7,30.3,-46.7) & = \\
velocity [km/s] & (26.9,-23.3,71.5) & $\times 0.5$ \\
spin axis & (0.65,0.65,-0.40) & =\\
\hline
\end{tabular}\\
``N'' stands for Newton and ``M'' for MOND. ``A'' indicates the orbit of the Antennae model of \citet{Renaud2015a}, and ``S'' represents lower initial velocities (``Slow'').
\end{table}

The initial conditions of the interactions are listed in \tab{ic_orbit}. The angles between the spin axis of the galaxies and the vector normal to the orbital plane are $46^{\circ}$ and $58^{\circ}$ for the progenitors 1 and 2 respectively, meaning that the interaction is prograde (i.e. with spin inclination $< 90^{\circ}$) for both galaxies, but with a spin-orbit coupling slightly stronger for galaxy 1 than for galaxy 2. (Prograde encounters favour the formation of long tidal features, see e.g. \citealt{Duc2013}.)

The progenitor galaxies are initially placed at a large distance from their encounter position ($\sim 60 \kpc$ each). This ensures that, for the velocities adopted, (i) they start in a quasi-isolation stage (in agreement with the initial setup of the models), (ii) they have enough time to virialise and evacuate the imperfections of the initial conditions before the interaction itself, (iii) they interact before the formation of substructures (bar, spirals) in the discs.

We also run Newton and MOND simulations with an initial velocity 1.75 times larger than \newta and \monda (not shown in this paper for the sake of clarity). These complementary simulations lead to the same conclusions as those presented below.

%%%%%%%%%%%
\subsection{Comparing dark matter and MOND simulations}
\label{sec:pb}

Ideally, to allow for direct comparisons, only the equation of gravitational acceleration and the presence of a DM halo should change between the Newton and MOND runs. In practice, things are more complicated.

First, the galaxy models must be setup in equilibrium. The velocities of the baryonic components must be set according to the local gravitational potential. Although MOND provides a good fit to the Newtonian rotation curve of galaxies (with an isothermal DM halo) at large distances, differences can be found in the inner regions ($\lesssim 10 \kpc$), depending on the shape of the DM halo MOND replaces. Setting up stable galaxies would then require to adjust the velocity dispersions of baryons between the two cases, and thus would lead to slightly different galaxies. Such difference could amplify over a few rotation periods and significantly alter the formation of sub-structures like bar(s), spiral arms and clumps. In turn, this would change the morphology of the galaxies, their intrinsic star formation activity and their response to the interaction. We circumvent this problem by using a specific galaxy model allowing for replacing the DM halo with the MOND formalism and maintaining equilibrium, as described in \sect{ic}.

Second, the use of an adaptive grid can also introduce a bias. In \ramses and \por, the refinement of the grid is based (among other criteria) on the number of particles in cells. Typically, a cell is refined when it contains more than 40 particles (but this number varies from simulation to simulation). Together with refinement criteria on mass and stability, this ensures that the baryonic component and the DM halo are correctly sampled on the AMR grid. However, in the MOND case, the absence of DM particles implies that the net number of particles per cell is never exactly the same as in the Newtonian run. This leads to different refinements or in other words, to different resolutions between the Newton and the MOND cases. However, the refinement is usually dictated by the total mass (stars, gas and DM) and the stability (based on the local Jeans length), both dominated by the baryons in the galactic discs. For this reason, and to avoid introducing numerical artefacts in changing the refinements criteria, we chose to keep the same strategy in the MOND and Newton cases. In practice, the differences in refinement are small and their effects on the physical state of the galaxies are negligible, as shown in \app{refinement}.

%%%%%%%%%%%
\subsection{Initial conditions}
\label{sec:ic}

\begin{table}
\caption{Initial setup of the progenitors}
\label{tab:ic}
\begin{tabular}{lc}
\hline
\multicolumn{2}{l}{Gas disc (exponential)}\\
mass [$\times 10^9 \msun$] & 4.2\\
characteristic radius [kpc] & 4.5\\
truncation radius [kpc] & 22.5\\
characteristic height [kpc] & 0.45\\
truncation height [kpc] & 2.25\\
\hline
\multicolumn{2}{l}{Stellar disc (exponential)}\\
number of particles & 720,000\\
mass [$\times 10^9 \msun$] & 41.62\\
characteristic radius [kpc] & 4.5\\
truncation radius [kpc] & 22.5\\
characteristic height [kpc] & 0.45\\
truncation height [kpc] & 2.25\\
\hline
\multicolumn{2}{l}{Stellar bulge (King)}\\
number of particles & 400,000\\
mass [$\times 10^9 \msun$] & 23.1\\
radial extent [kpc] & 4.5\\
\hline
\multicolumn{2}{l}{Dark matter halo (lowered Evans)}\\
number of particles & 600,000\\
mass [$\times 10^9 \msun$] & 261.0\\
concentration & 0.1\\
characteristic radius [kpc] & 3.6\\
truncation radius [kpc] & 50.0\\
\hline
Total mass [$\times 10^9 \msun$] & 330.0 (Newton), 69.0 (MOND)\\
\hline
\end{tabular}
\end{table}

In their gas-free study, \citet{Candlish2015} found that the composite disc-bulge-halo model MW-A from \citet{Kuijken1995} yields a very similar behaviour over several Gyr when the DM halo is replaced with QUMOND phantom dark matter. Therefore, such a model minimizes the differences between the MOND and Newton cases and solves the first point of \sect{pb}. For this reason, we decide to use this model in all our simulations\footnote{The main differences with the models of \citet{Renaud2015a} are a larger disc and a more massive bulge. These changes affect the evolution and the properties of the merger, in particular the SFR, as discussed in \sect{sfr}.}. The $N$-body components (stars and DM) are generated using the \texttt{mkkd95} tool from the \nemo package \citep{Teuben1995}. Then, particles representing $10\%$ of the stellar disc mass are removed and replaced by the equivalent density distribution in gas form (in \ramses cells). A rotation velocity is attributed to the gas cells to balance the local gravitational, turbulent and sonic pressure terms (see \citealt{Chapon2011}, \citealt{Renaud2013b}). Within a dynamical time, the gas component cools down into a thinner disc. Such an approach simplifies the initial setup in \ramses, and is followed for both our Newton and MOND runs. In the MOND case, the DM particles are deleted. The initial conditions of our models are summarized in \tab{ic} (see also MW-A from \citealt{Kuijken1995}).

To verify the stability of our setups, we have run simulations of the progenitor galaxy in isolation, both in the Newton and MOND framework. Results are summarized in \app{isolated}.

%%%%%%%%%%%
\subsection{Extended halos}

As shown by \citet{Bournaud2003}, the formation of structures in tidal debris is sensitive to the size of the DM halo. Extended halos can indeed favour the local collapse of gas and the further formation of stellar objects like clusters and TDGs. Since the MW-A model we use has initially been designed to study structures in isolated disc galaxies, the halo is truncated to a rather small value compared to typical virial radii ($50 \kpc$ instead of $\approx 200 \kpc$ for the Milky Way, see e.g. \citealt{Dehnen2006}). This provides a significant gain in computational time and memory. In studies like ours however, the tidal tails of interacting galaxies could expand to large distances and our results could thus be affected by the artificial truncation of the halo. (In practice however, the tidal tails of our model are well within the truncation radius, see \sect{morpho}.)

To test the importance of the truncation, we replace the truncated lowered Evans DM halo of our models with a NFW \citep{Navarro1997} model parametrized such that it reproduces the velocity curve in the innermost $30 \kpc$, and truncated at much larger radius ($200 \kpc$). Details are presented in \app{extended}. Without changing the baryonic components of our models, we run the \newta case with the extended halo. Apart from slight deviations due to the different mass of the progenitor galaxy, we find no difference in the global morphology nor in the substructures of the tidal tails.

We conclude that the truncation radius adopted here ($50 \kpc$) is sufficiently large not to affect the formation of tidal tails nor substructures inside them. We decide to keep the truncated version (as in \tab{ic}) to minimize the computational cost of the simulations.

%%%%%%%%%%%%%%%%%%%%%%%%%%%%%%%%%%%%%%%%%%%%%%%%%%%%%%%%%%%%%%%%%%%%%%%%%%%%%%%%
\section{Results}

%%%%%%%%%%%
\subsection{Orbits}

\begin{figure}
\includegraphics{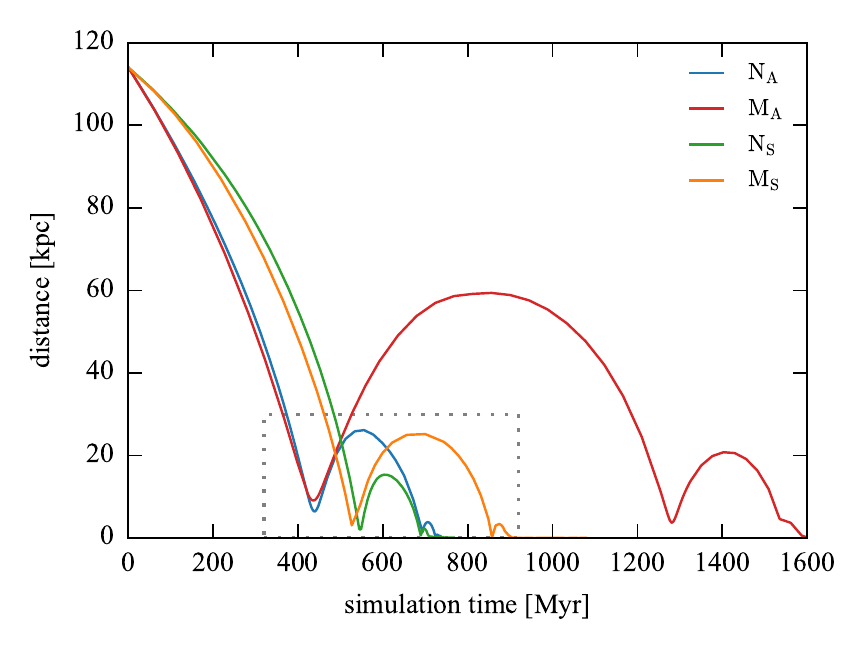}
\includegraphics{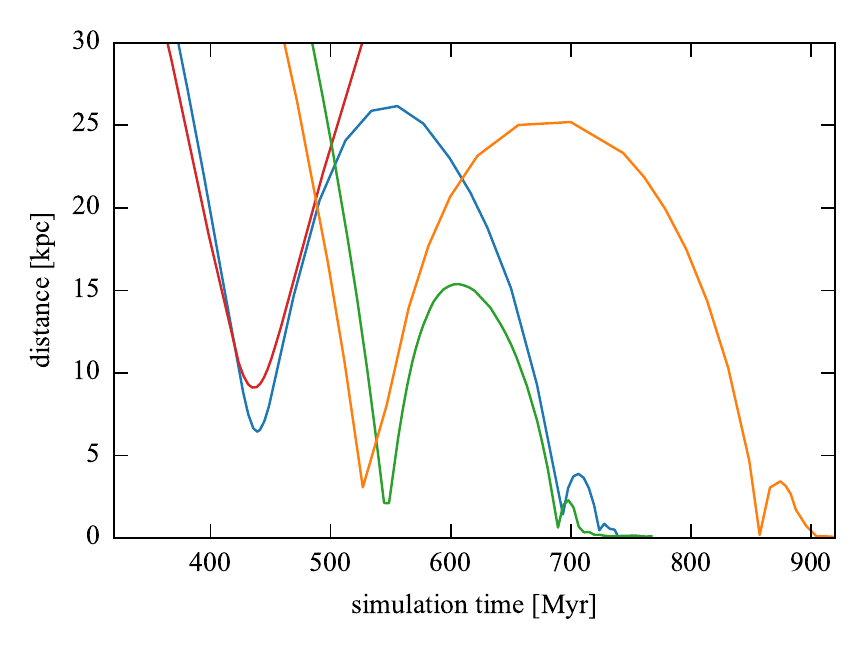}
\caption{Evolution of the distance between the progenitor galaxies, starting at the beginning of the simulations. The bottom panel is a zoom-in. All models start with a separation of $114 \kpc$ at $t=0$.}
\label{fig:distance}
\end{figure}

In MOND, as galaxies come close to each other, they still lie in the weak gravitational regime where MOND provides the strongest differences from the $\Lambda$-CDM framework. The change in dynamical friction due to the absence of a DM halo significantly alters the exchange of orbital energy and thus the fate of the progenitors. For that reason, most of interacting galaxies in MOND have to be fly-bys and the frequency of galaxies actually merging is much lower in MOND than in Newtonian dynamics \citep{Nipoti2008, Kroupa2015}. For the same reason, it is already known that when the orbit allows for it, the merger of galaxies in MOND takes longer than in $\Lambda$-CDM \citep{Tiret2008b}. Furthermore, because of the absence of DM, all exchanges of orbital angular momentum are exclusively transmitted to the baryons, which also affects the disruption of the galaxies at the moment of their encounter(s) \citep{Combes2010}.

\fig{distance} shows the evolution of distance between the two galaxies for each simulation. We compute this as the distance between the centres of mass of the $\approx 1000$ initially most gravitationally bound stellar particles of the galaxies. All simulations have been stopped after final coalescence, once the global SFR decreases back to a few $\Msunyr$.

\begin{table}
\caption{Orbits}
\label{tab:orbits}
\begin{tabular}{lcccc}
\hline
Simulation & \newta & \monda & \newts & \monds \\
\hline
First pericentre [kpc] & 6.5 & 9.2 & 2.0 & 3.1 \\
Second pericentre [kpc] & 1.4 & 3.8 & 0.6 & 0.2 \\
Separation phase$^{(1)}$ [Myr] & 255 & 845 & 143 & 331 \\
Interaction phase$^{(2)}$ [Myr] & 284 & 1149 & 165 & 368 \\
Coalescence [normalised time$^{\star}$] & 1.11 & 1.36 & 1.15 & 1.11 \\
\hline
\end{tabular}\\
$^{(1)}$: time between the first and second passages.\\
$^{(2)}$: time between the first passage and the coalescence.\\
$^{\star}$: in normalised time, the first (resp. second) pericentre passage occurs at $t' = 0$ (resp. $1$). Normalised time at coalescence is thus the ratio of the durations of the interaction and the separation phases.
\end{table}

The differences between the Newton and MOND cases arise a few Myr before the first pericentre passage, due to the differences in dynamical friction from the DM component of the progenitors when they start to overlap. Because of the small impact parameters of the orbits used, the baryonic components also overlap adding an extra contribution to dynamical friction. In the MOND cases, only this baryonic contribution acts on the braking of the galaxies (on top of the transfer of kinetic energy and angular momentum to the tidal debris).

Because of the diversity of orbits arising from the initial velocities we choose and the differences in dynamical friction, comparing the evolution of physical quantities in our simulations can be difficult. We circumvent this by defining the time $t$ with respect to the first pericentre passage, in the rest of the paper. Furthermore, we introduce the normalised time of interactions obtained by normalising the time to the duration of the separation phase, i.e. the period between the first and second pericentre passages. Therefore in the following, a normalized time $t'=0$ corresponds to the first passage, while $t'= 1$ corresponds to the second passage.

The pericentre distances and the time spans between encounters are listed in \tab{orbits}.

%%%%%%%%%%%
\subsection{Morphology}
\label{sec:morpho}

\begin{figure*}
\includegraphics{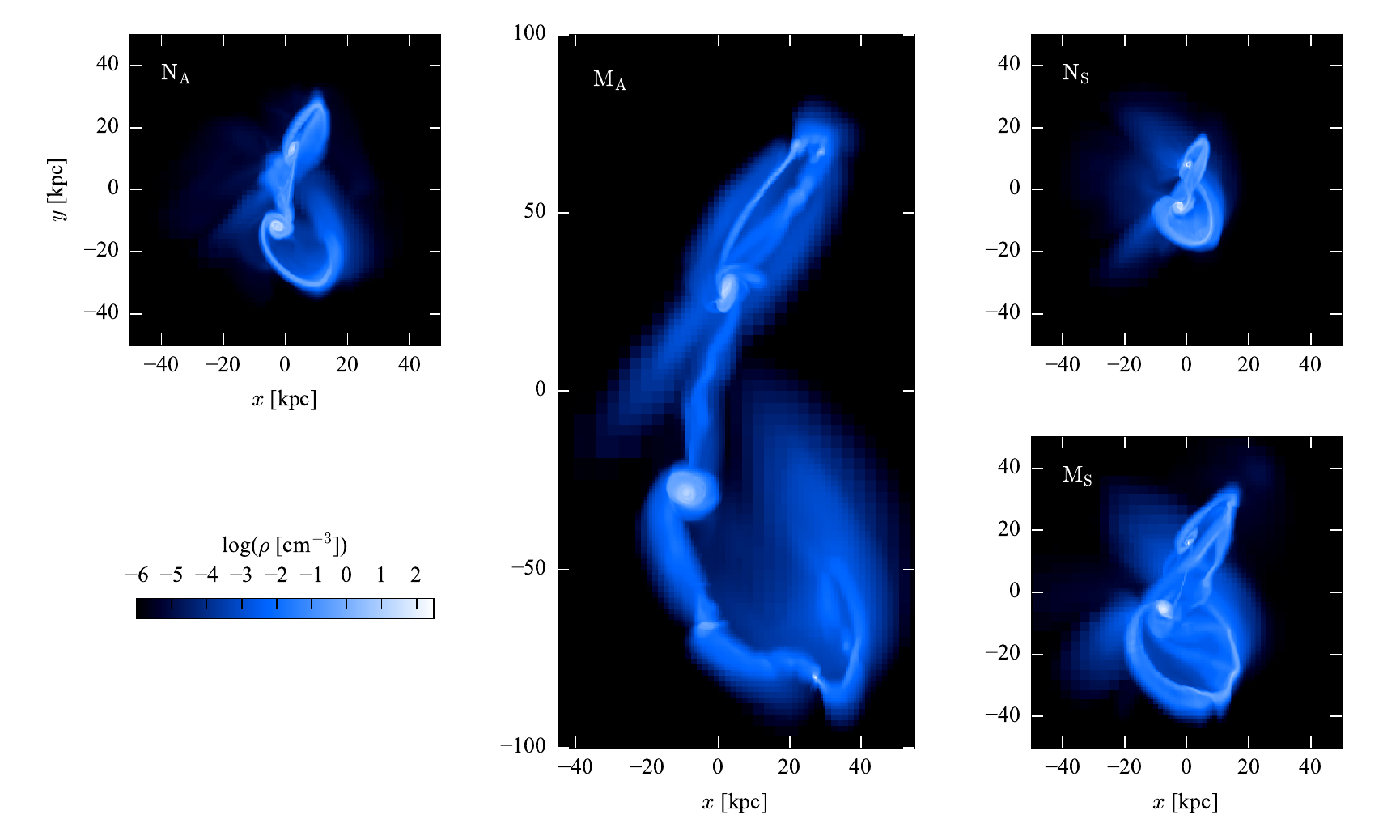}
\caption{Map of the gas density of the models, when the galaxies reach their maximum separation after the first passage. To ease the comparison of the spatial extent of the tidal tails, all panels are at the same scale. Note the sharper and more structured features due to the baryons self-gravity in the Milgromian dynamics, compared to the Newtonian one.}
\label{fig:morpho}
\end{figure*}

\begin{figure*}
\includegraphics{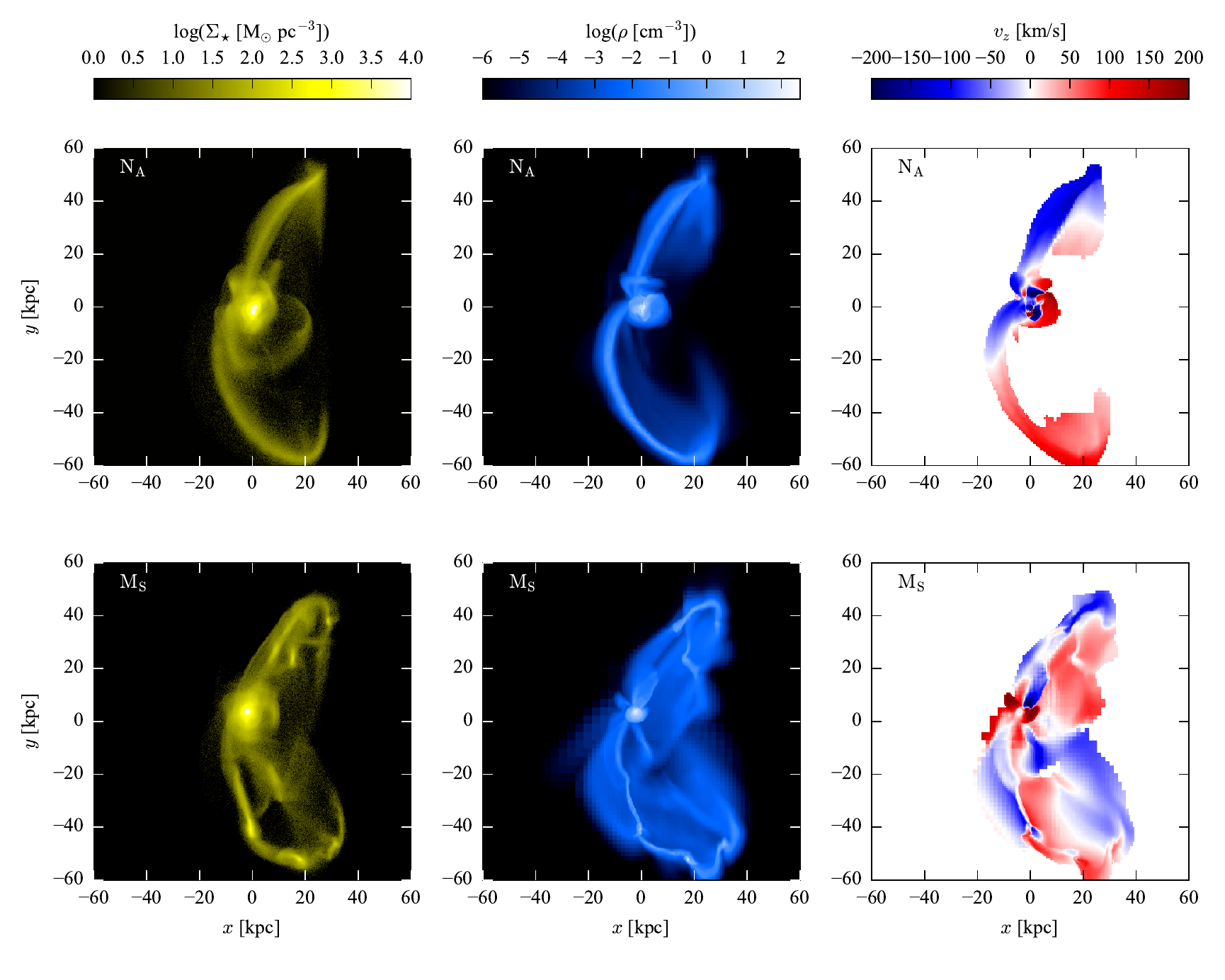}
\caption{Map of the stellar surface density (left), gas density (centre) and velocity along the line of sight (right) of the \newta (top) and \monds (bottom), at the second pericentre passage. For the sake of clarity, only the velocity field in regions of high gas density is shown.}
\label{fig:morphoant}
\end{figure*}

\fig{morpho} displays the morphologies of the models at the epochs of maximum separation between the first and second passages. In addition, \fig{morphoant} shows the stellar and gas density maps, and the velocity along the line of sight at the moment of the second pericentre passage\footnote{The real Antennae system is thought to be observed a few Myr before the second pericentre passage, when the discs start to overlap on the Eastern side, and before nuclei reach their minimum distance \citep{Renaud2008}.}.

In the \monda model, the high relative velocity between the progenitors after the first passage (due to the initially high velocity and the weak effect of dynamical friction) accounts for a larger separation, and a longer period for the tidal tails to expand. At the moment of maximum separation between the progenitors, the two most distant points in the stellar tidal tails are separated by $\approx 230 \kpc$ ($\approx 160 \kpc$ in the plane of the sky), i.e. 3.4 times more than those of \newta (2.7 times in the plane of the sky).

Whereas the progenitors are initially setup with a relative velocity twice higher in \newta than in \monds, because of the differences in dynamical friction they reach similar velocities at their respective pericentre passages ($594 \kms$ and $568 \kms$ respectively). That is, both models have comparable relative velocities, impact parameters and masses at the time of the first encounter, which translates in the expected comparable morphology and spatial extent of the tidal tails. After the first passage, differences in dynamical friction continue to alter the evolutions of the two models and make the separation phase of \monds $\approx 1.3$ times ($76 \Myr$) longer than that of \newta. Yet, because of differences in angular momentum transfer, the lengths of the tidal tails remain comparable in both formalisms.

We note that the disc remnant of one of the galaxies in the \monds model (Galaxy 1, located at $y > 0$ during the separation in the bottom-right panel of \fig{morpho}) is more concentrated than its equivalent in \newta. This galaxy is that having the strongest spin-orbit coupling (or equivalently the lowest inclination in the orbital plane, see \sect{suite}) and thus the differences between the two formalisms and with the other galaxy are likely due to different responses in the angular momentum transfer during the first passage. This further affects the evolution of the system at coalescence (see below).

In \monds, the tidal tails host the formation of over-densities similar to tidal dwarf galaxies (about 3 in each tail are visible in \fig{morphoant}). Their associated phantom dark matter further favours their growth, while no such structure is visible in the stellar nor gaseous component of the \newta model. The formation of TDGs in simulations is sensitive to the resolution \citep{Wetzstein2007} and the truncation of the DM halo \citep{Bournaud2003}. By conducting our comparisons at the same resolution in Newton and MOND, and by testing the formation of sub-structures with much more extended halos (\app{extended}), we ensure that the differences we detect have a physical origin. Observations of the Antennae galaxies report only one TDG candidate, at the tip of the Southern tail \citep{Mirabel1992}, but the exact nature of this structure is still questioned \citep[see also \citealt{Bournaud2004}]{Hibbard2001}. It could be either an unbound object or a forming TDG still out of equilibrium. For the models we consider, the Newtonian framework does not allow for the formation of TDGs, while the Milgromian dynamics does. However we note that, in the absence of efficient shielding from the rest of the galaxy, star forming regions in the tidal tails are more sensitive to ultraviolet radiation of extragalactic origin. A stronger radiation, e.g. at higher redshift or in a denser galactic environment, could potentially prevent the formation of TDGs and thus reduce the differences (in the young stellar component) between Newtonian and Milgromian cases in this context. Our simulations show however that the old stellar component is likely to remain more clumpy in MOND, as long as the gaseous contribution to the local gravitational potential is negligible over the stellar one. Leaving this issue aside, with the specific models considered here, the Milgromian runs tend to slightly overproduce TDGs given the absence of unambiguously defined ones in the Antennae, whilst the Newtonian runs might potentially slightly underproduce them if the observed TDG candidate turns out to be real. However, the uniqueness of our initial conditions has not been established and it is possible that other sets of parameters could reproduce the same morphology with a different number of TDGs. A much larger simulation sample including more interacting systems would be necessary to reach a clear conclusion on this particular topic.

In conclusion, despite different small scale features ($\lesssim 1 \kpc$) and without a fine tuning of the parameters, the overall morphology and kinematics of an Antennae-like system can be reproduced in the Milgromian dynamics framework when the progenitors have a small initial velocity to compensate for the weakness of dynamical friction. This complements the pioneer work of \citet{Tiret2008b} who reproduced an Antennae-like morphology in MOND\footnote{Their model uses the same orbit as that of the restricted simulation of \citet{Toomre1972} with point-mass galaxies surrounded by mass-less tracer particles, thus with no dynamical friction.}.

The main differences between the Newton and MOND Antennae-like models are the concentration of the disc remnants, and the presence of sub-structures along the tidal tails. Both these features play an important role in the star formation histories of the mergers, as discussed below.

%%%%%%%%%%%
\subsection{Star formation}
\label{sec:sfr}

The starburst activity of interacting galaxies can be triggered by several physical processes, each having different properties and signatures.
\begin{itemize}
\item Inflow of gas toward the galactic centres \citep[see e.g.][]{Barnes1991, Hopkins2009, Bournaud2010a}. By definition, such activity is restricted to the innermost regions of the galaxies and is mostly triggered during close encounters, when one galaxy exerts negative gravitational torques on the ISM of the other, inside co-rotation. These torques fuel the gas inward and thus increase its density in the innermost regions of the galaxies. This effect can also be amplified by torques generated internally by the disc structures like interaction-triggered bars and spiral arms.
\item Collisions between clouds and \hi reservoirs \citep{Jog1992, Barnes2004}. When the ISM of the progenitors overlap, it is possible that shocks make previously marginally stable gaseous structures collapse. Such an effect is limited to the overlap phase(s) and region(s) of penetrating encounters and is thus mostly active and efficient at the late stages of interactions like at coalescence. It is also possible that the overlap volume hosts cloud-cloud collisions, known to trigger intense episodes of star formation, in particular in the form of massive stars \citep[see e.g.][]{Loren1976, Tan2000, Renaud2015d}.
\item Compression of the ISM by tides and turbulence \citep{Renaud2014b}. The gravitational interaction between galaxies make the tidal field fully compressive over kpc-scale volumes \citep{Renaud2008, Renaud2009}. This effect is transmitted to the turbulence which develop a compressive nature, resulting in excesses of dense gas and thus an elevation of the SFR \citep[see also][on stability criteria]{Jog2013, Jog2014, Mondal2015}. This process is active during all passages and over large volumes.
\end{itemize}
Therefore, central star formation activity is more likely to be linked to inflows. Activity in the overlap regions of the progenitors can be associated to collisions and/or compression. Finally, compression is likely responsible for triggering starburst activity over large volumes, in particular in the outskirts of galaxies and the tidal debris. In all cases and all regions, star formation can obviously still proceed in a non-enhanced way, and/or with small scale triggers, as it does in isolated galaxies.

%%%%%%%%%%%
\subsubsection{Global evolution}

\begin{figure*}
\includegraphics{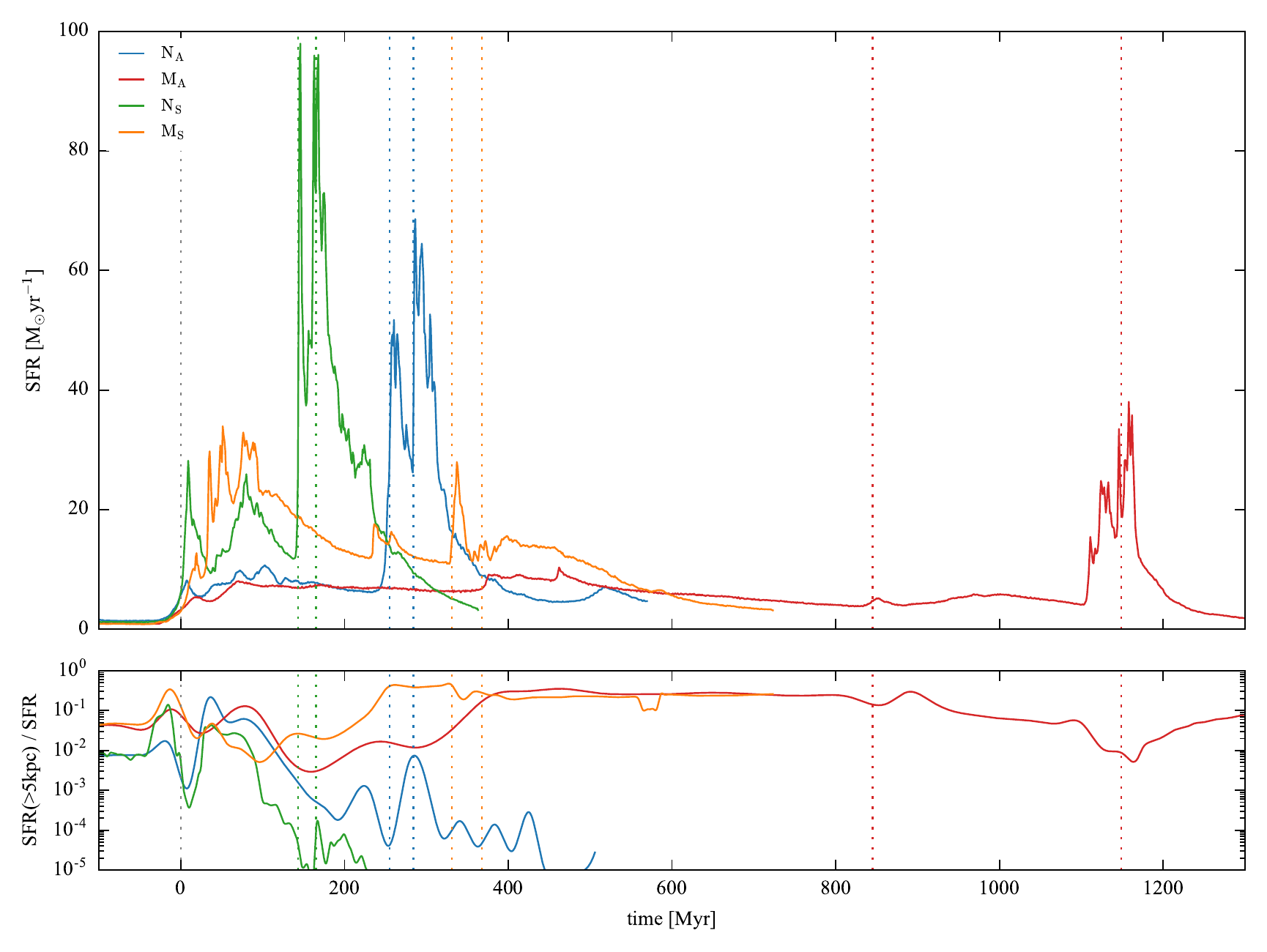}
\caption{Star formation rate (top), and fraction of the SFR further than $5 \kpc$ from the galaxy centres (bottom). A large fraction denotes star formation activity in the tidal debris. $t=0$ corresponds to the first pericentre passage. Vertical lines indicate the first and second pericentre passages, and the final coalescence.}
\label{fig:sfr_center}
\end{figure*}

\begin{figure}
\includegraphics{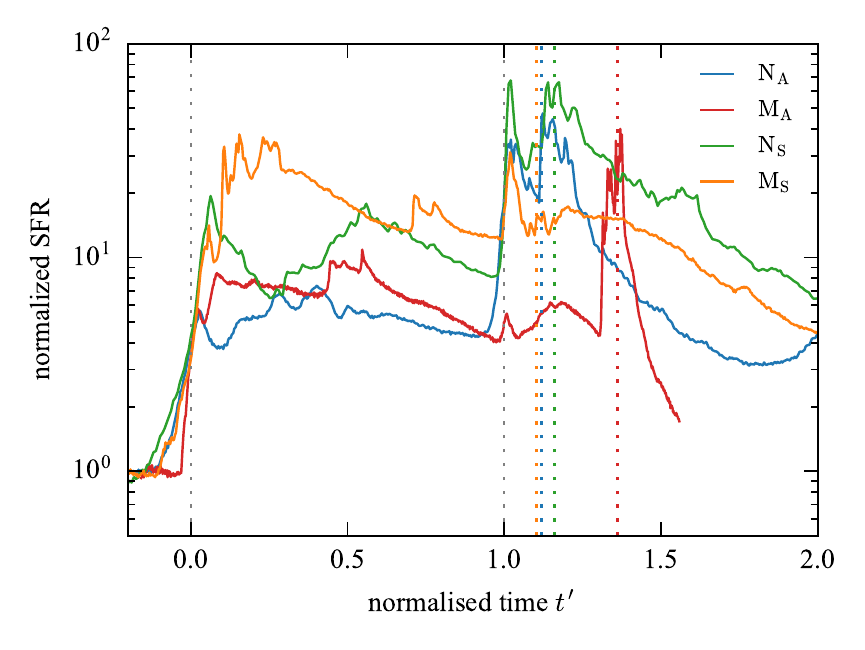}
\caption{Star formation rate normalized to its value before the interaction, as a function of time normalized to the separation period (i.e. 0 on the x-axis corresponds to first pericentre passage, while 1 indicates the second passage). The vertical lines indicate final coalescence, corresponding to the third encounter for these models.}
\label{fig:sfr_norm}
\end{figure}

\fig[s]{sfr_center} and \ref{fig:sfr_norm} show the star formation rates of all models. The differences in the ISM noticed for the isolated models (see \app{isolated}) induces slightly different intrinsic SFRs. For our Newtonian models, the total SFR in the isolation phase (i.e. before any sign of interaction) is $1.4 \Msunyr$, while it is $0.9 \Msunyr$ in the MOND cases. We also note that the MOND models have a slightly more spatially extended star formation in isolation: $3\%$ of their stars form beyond $5 \kpc$, while this fraction is less than $1\%$ in the Newtonian cases.

All simulations yield an increase of their star formation activity at the times of all their pericentre passages, but with remarkably different amplitudes. Although the relative importance of the physical triggers responsible for these bursts varies from one model to the next, any enhancement of the star formation activity is stronger when its trigger acts for a long period. Therefore, systems with low relative velocities (\newts and, to a lower extent \monds) harbour a stronger burst of star formation. The fastest model (i.e. with the highest initial velocity and the least dynamical friction) \monda yields the lowest burst at the first passage. In short, the intensity of the first burst in our models decreases with increasing velocity of the progenitors. This result seems to contradict the trend proposed by \citet[their Figure~22]{DiMatteo2007}, but we note that their sample of direct fly-bys for late-type galaxies (which corresponds to all our cases at first passage) does not yield a clear trend between the intensity of starburst and the relative velocity of the progenitors in the range covered by our models ($\approx 500 \mh 600 \kms$ at pericentre).

In all models, triggered star formation starts a few Myr before the pericentre passage itself, and over larger volumes than during the isolation phase. This is likely the signature of the long range effect of tidal compression. Just after the pericentre, the fraction of SFR in the outer regions drops, indicating that gas has been fuelled toward the nuclei and forms stars. The delay between the trigger (torques) and the resulting activity (SFR enhancement) is slightly shorter in the Newton runs than in our Milgromian models.

After the pericentre passage, as the galaxies start to separate, the tidal debris is ejected. The presence of a massive bulge in our models helps maintaining a relatively high SFR during the separation phase, while it rapidly decreases down to almost its pre-interaction value in the simulation of \citet{Renaud2015a} who used the same orbit as \newta but different progenitor galaxies. Interestingly, the maximum SFR before coalescence corresponds to the pericentre passage for the Newton models, while it is reached later during the separation in the Milgromian runs, and is associated with formation outside the disc remnants, i.e. in the weak gravitational regime, exactly where differences between Milgromian and Newtonian dynamics are expected (see \sect{outer}).

At the time of the second passage and coalescence, the configuration of the first passage repeats itself. Because of the orbital energy transferred to the tidal debris, the impact parameter and relative velocities are however smaller than before (see \fig{distance}) and the above-mentioned effects (torques, collisions in overlaps) get even stronger than at the first encounter. Thus, most of the star formation takes place in the central regions. In the \monda model, the second passage is very similar to the first one, both having relative velocities of $\approx 500 \kms$, leading to a weak enhancement of the SFR (to about $5 \Msunyr$).

At coalescence, all models host a burst of star formation activity, which lasts for about $\approx 100 \Myr$. The intensity of this burst depends, as before, on the orbital configuration, but also on the amount of gas left in the reservoir after the precedent bursts.

%%%%%%%%%%%
\subsubsection{Star formation in the tidal debris}
\label{sec:outer}

For all models, enhanced star formation activity starts a few Myr before the first pericentre passage, in the outer regions. This corresponds to the distant effect of compressive tides acting on large scales over the progenitors. The tidal debris, namely a bridge connecting the two progenitors and tails on the other sides, forms and expand during the separation phase. In the Newtonian cases the gas clouds in the tidal debris are destroyed by dynamical processes like extensive (classical, destructive) tides and shear \citep{Renaud2015a}. Star formation in the outer regions is thus quenched.

The ISM of the tidal bridges connecting the two galaxies is more fragmented than the tails (\fig{morpho}), and hosts the formation of a few star clusters (see also \citealt{Renaud2015a}). There, the inner regions of the two DM halos overlap, as opposed to the tails where the more distant halo of the companion has a negligible effect. 

In the MOND cases however, the tidal debris generates its own ``phantom'' potential well which favours the gathering of gaseous over-densities, leading to star formation in TDG-like objects (\fig[s]{morpho} and \ref{fig:morphoant}). These formation events are visible in the total star formation rate (\fig{sfr_center}) as peaks during the separation phase (e.g. at $t\approx 240 \Myr$ in \monds, and $t\approx 460 \Myr$ in \monda). At these moments, the fraction of SFR in the outer regions increases significantly (although the activity in the discs also continues). This constitutes a major difference between the Newtonian and Milgromian families of models. 

%%%%%%%%%%%
\subsubsection{Comparisons between Newtonian and Milgromian gravitations}

In this Section, we focus on the differences between our most comparable models, namely \newta and \monda which share the same initial conditions, and \newta and \monds which yield comparable morphologies.

\begin{figure}
\includegraphics{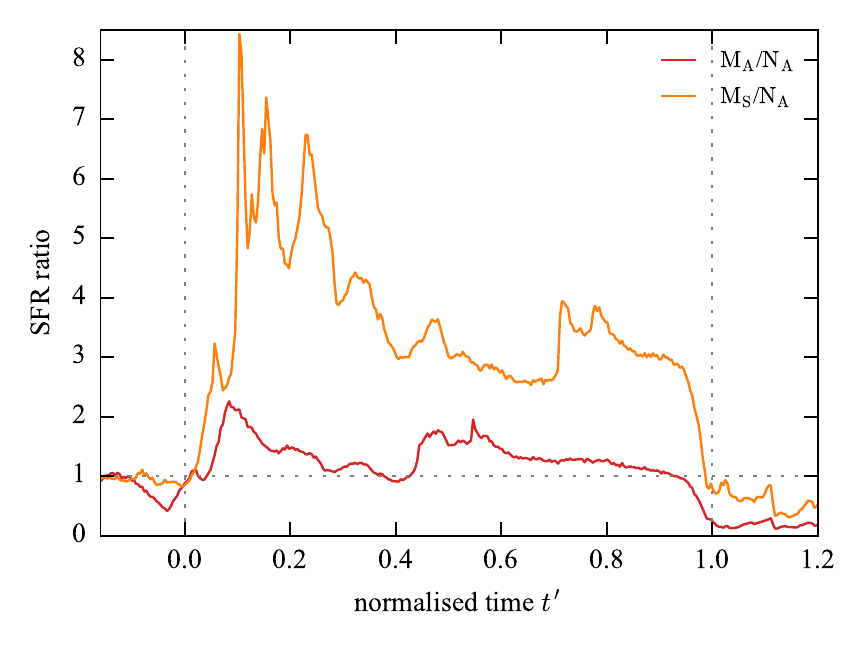}
\caption{Ratio of the SFRs between the \monda and \newta, and the \monds and \newta simulations, i.e. our two sets of most direct comparisons. As in \fig{sfr_norm}, time is normalized to the separation period. Note that, in this time unit, \newta, \monda and \monds models reach coalescence at 1.12, 1.37 and 1.13 respectively.}
\label{fig:sfr_mond_vs_newt}
\end{figure}

\fig{sfr_mond_vs_newt} shows the ratios of SFR between our models. The increase of the SFR at the pericentre passages is fairly independent of the gravitational paradigm (\newta and \monda) but varies with the velocity of the encounter (\newta and \monds). We note from \fig{sfr_norm} that the first stages in the star formation burst a few Myr before the pericentre itself ($-0.1 \lesssim t' \lesssim 0$), are remarkably similar in the \newta and \monds runs. Since this phase is mostly associated with tidal compression, we conclude that the Milgromian gravitation produces here a similar effect as the DM halo at these epochs. The same situation is found again at the moment of the second encounter passages ($t' \approx 1$), despite very different evolutions in between.

As noted before, galaxies in the Milgromian framework are more efficient at forming stars in between the pericentre passages than the Newtonian case. A significant fraction of this activity (up to $40 \%$) is located in the tidal debris, but the majority happens in the disc remnants. At this stage, star formation takes place in the fragmented ISM and because of tidal compression. However at large separations, the Newtonian galaxies lie in the outer regions of the DM halos of each other, where tidal compression (if/where/when it exists) is weaker than in the inner halo \citep{Renaud2009}. The Milgromian formalism is then more efficient at triggering star formation in the discs, in a comparable fashion as the situation in tidal debris (see \sect{outer}).

The Newtonian case is then forming more stars after the second passage ($t' > 1$) than both Milgromian models. This activity takes place in the disc remnants of the galaxies, in the central $\approx 5 \kpc$, as visible in \fig{sfr_center}. Gas inflows driven by gravitational torques favours such star formation. The differences between the Newtonian and Milgromian models come from the amount of gas available in the volume affected by these torques, which depends on the angular momentum transfer during the first passage (see the different disc morphologies and densities in \fig{morpho}). Even though torques of comparable amplitude exist in all models, the Newtonian one has retained more gas to fuel the nuclear star formation activity than the Milgromian cases. A similar effect can be noted in retrograde encounters where momentum transfer is very inefficient at the first passage, which maintains a large gas reservoir for the later encounters \citep[see e.g.][]{Duc2013}.

Because of the longer interaction period of the Milgromian runs, over the entire interaction (i.e. from $40 \Myr$ before the first passage until the end of the simulations, when the SFRs have approximately reached back their pre-interaction level), the total mass of stars formed is $6.7 \times 10^9 \Msun$ in the \newta run, and 30\%, 13\% and 37\% more in the \monda, \newts and \monds models respectively. However, the spread in formation epochs makes such differences difficult to detect observationally, as discussed in the next Section.

To conclude, significant differences exist in the star formation activities of the Newtonian and Milgromian models, in particular in between the pericentre passages. Part of these differences corresponds to star formation in the tidal debris, as noted in \fig{sfr_center}. Such activity does not exist in the Newtonian model, which is less prone to form TDGs in this simulated encounter as mentioned earlier.

%%%%%%%%%%%
\subsection{Observational diagnostics}

\begin{figure}
\includegraphics{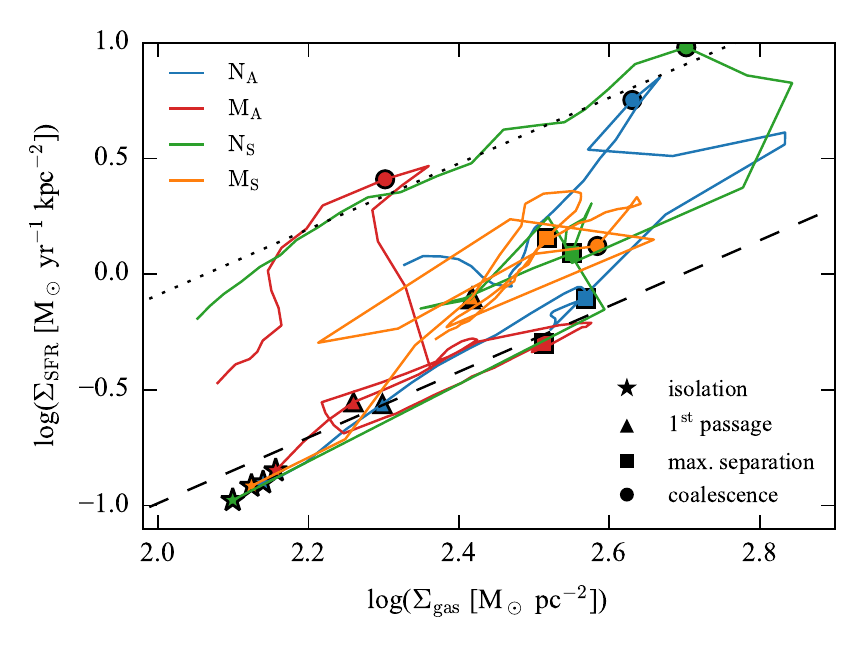}
\caption{Evolution of the gas and SFR surface densities in the central $2 \kpc$ of the galaxy(ies). Before coalescence, the points are the average of the two progenitors. The dashed and dotted lines indicate the observed sequences of discs and of starbursts respectively, as in \citet[see also \citealt{Genzel2010}]{Daddi2010a}. Stars, triangles, squares and dots mark the moments before the interaction, during the first starburst (i.e. a few Myr after the first passage), at the maximum separation of the galaxies, and during the second starburst (about at the coalescence).}
\label{fig:ks}
\end{figure}

\fig{ks} shows the evolution of the models in the Schmidt-Kennicutt diagram, linking the surface density of gas $\Sigma_\mathrm{gas}$ to the surface density of SFR $\Sigma_\mathrm{SFR}$. By construction, all models start close to the regime of discs noted by \citet{Daddi2010a} and \citet{Genzel2010}. At the first passage, they become more compact and more efficient at forming stars, as discussed above. The maximum efficiencies are reached around the coalescence. The Newtonian models reach higher densities than the MOND ones, again because of differences in angular momentum transfer away from the central regions.

\begin{figure}
\includegraphics{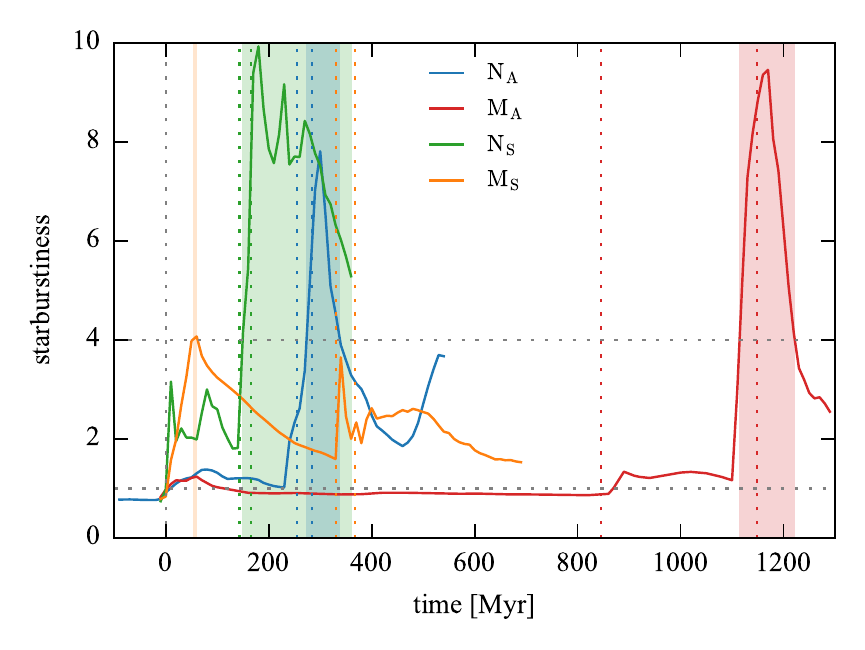}
\caption{Evolution of the starburstiness parameter, defined as the ratio between the measured $\Sigma_\mathrm{SFR}$ and that of the disc sequence for the corresponding measured $\Sigma_\mathrm{gas}$. $t=0$ corresponds to the first pericentre passage. The shaded areas mark the star-bursting phases of the interactions (i.e. with a starburstiness above 4).}
\label{fig:starburstiness}
\end{figure}

Following Fensch et al. (in prep), we define the starburstiness parameter as the ratio of the measured surface density of SFR ($\Sigma_\mathrm{SFR}$) and the value it would have if the galaxies were on the disc sequence for the measured surface density of gas ($\Sigma_\mathrm{gas}$). We consider the galaxies as star-bursting when their starburstiness exceeds 4 \citep[see e.g.][]{Schreiber2015}. \fig{starburstiness} shows the evolution of the starburstiness for the four models. Galaxies are considered as star-bursting for $65$, $108$ and $9 \Myr$ for the \newta, \monda and \monds runs respectively, and for more than $213 \Myr$ in the \newts run where we stopped the simulation before the system leaves this regime. Since the starburst regime is only reached (if it is at all) at the pericenters and not during the entire interaction, the longer interaction periods of the MOND cases do not imply a longer duration of the starburst phase. Therefore, for comparable morphologies, the Newtonian systems are more likely to be observed in a starburst phase like (ultra-)luminous infrared galaxies (LIRGs, ULIRGs, e.g. \citealt{Kennicutt1998b}) than their Milgromian equivalent.

\begin{figure}
\includegraphics{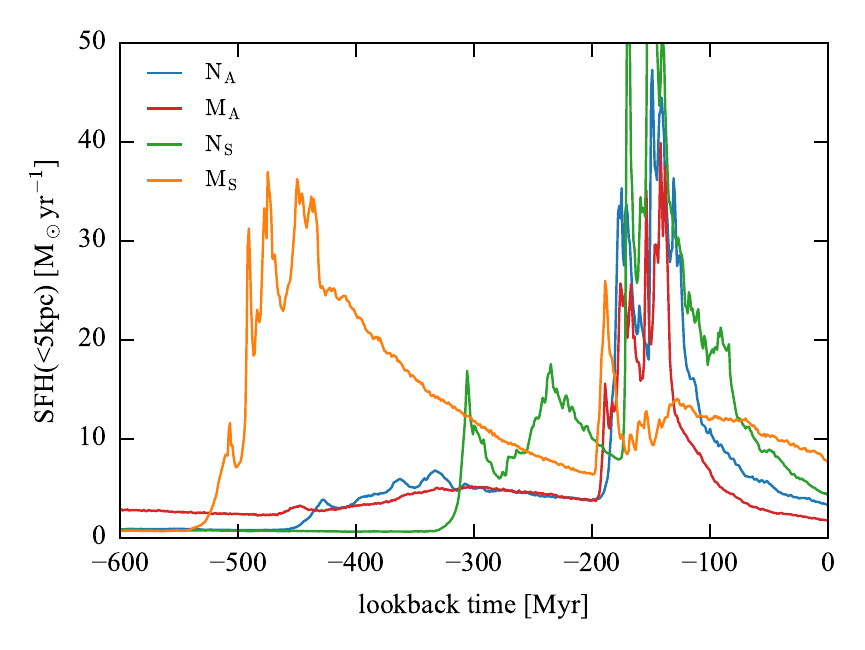}
\caption{Star formation history of the stars in the central $5 \kpc$. To adopt an observational perspective, here the time axis is reversed, and $t=0$ corresponds to $150\Myr$ after the beginning of final coalescence.}
\label{fig:sfh_center}
\end{figure}
 
Finally, from an observational perspective, \fig{sfh_center} shows the star formation history (SFH) of the merger remnants, i.e. the histogram of ages for the stars detected in the central $5 \kpc$, $150 \Myr$ after the beginning of final coalescence. As in observations, most of the stars form in this volume, but some have been accreted from the tidal debris. The recent star formation history (i.e. during coalescence, less than $200 \Myr$ in the past) is comparable for all models with differences in the amplitude of the burst as discussed above. However, the differences in orbital period and in the burst associated with the first encounter provide a strong dichotomy between the Newtonian and Milgromian models. When comparing models with similar morphologies (\newta and \monds), we note that the SFH in MOND is more extended and yields a stronger burst in the $\approx 300 \mh 600 \Myr$ range. 

To conclude, interacting galaxies in MOND host extended star formation in time and space, which makes them less likely to be detected as starbursts as their Newton equivalents.

%%%%%%%%%%%%%%%%%%%%%%%%%%%%%%%%%%%%%%%%%%%%%%%%%%%%%%%%%%%%%%%%%%%%%%%%%%%%%%%%
\section{Summary and discussion}

Using hydrodynamical simulations, we present a comparison as directly as possible between interacting galaxies within the Newtonian and MOND frameworks. Our main results are:
\begin{itemize}
\item Replacing dark matter with a MOND formalism induces differences in dynamical friction and angular momentum transfer during the galactic encounters. As a result, the tidal debris spans a much larger volume in Milgromian models than in Newtonian ones for a given set of initial conditions.
\item Comparable merger morphologies can still be obtained in Newtonian and Milgromian frameworks. We find that merely tuning down the relative velocity of the progenitors to balance the weak dynamical friction in MOND and reach comparable velocities at the first pericentre in both frameworks provides a reasonable correspondence between the models, at least on large scales. Thus, without a fine tuning of the parameters, the overall morphology and kinematics of an Antennae-like system can be reproduced in the MOND framework. This complements the pioneer work of \citet{Tiret2008b} who reproduced an Antennae-like morphology in MOND for the first time.
\item While the global morphology and kinematic structure can be reproduced with Milgromian dynamics, small scale differences from the Newtonian model exist: mainly, the formation of stellar and gaseous clumps along the tidal tails and the spatial extent of the discs at the moment of the second encounter. By generating their own phantom dark matter and thus a deep potential well on small scales ($\lesssim 1 \kpc$), the tidal tails favour the collapse of dense gas structures and thus the formation of stars. A significant fraction of the total star formation occurs in the tidal debris in Milgromiam gravitation, while such activity is negligible in the Newtonian case. The Milgromian models thus lead to significantly more sub-structures in the tidal tails than the Newtonian models.
\item The resulting star formation activity is thus significantly more extended in space and time in Milgromian than in Newtonian gravity.
\end{itemize}

The results presented here originate from a handful of orbital configurations and for only one galaxy model. Generalizing our conclusions to other systems, over the broad range of observed parameters (mass, mass ratio, relative velocity, inclination, spin-orbit coupling etc.) would require a more complete survey of simulations. Such a study is proposed in the Newtonian framework by the {\tt GALMER} project \citep{DiMatteo2007}, where hundreds of configurations are explored. An equivalent in Milgromian gravitation would allow us to broaden our conclusions on the response of Milgromian galaxies to interactions and mergers.

In particular, we expect that the SFH of retrograde encounters (i.e. orbital and disc angular momenta being anti-aligned) in Milgromian dynamics would be more similar to their Newtonian counterparts than for the cases we presented here. During the first pericentre passage of a retrograde encounter, the orbital angular momentum is inefficiently transferred to the discs and tidal features are less pronounced and shorter than in prograde cases \citep[see e.g.][]{Duc2013}. As a result, a large gas reservoir remains available at the late stages of the interaction, in particular at coalescence, where cloud-cloud collisions and nuclear inflows can efficiently trigger an intense episode of star formation. In that case, the SFH would only yield one major peak at coalescence. The main differences between Newtonian and Milgromian models noted above in term of space and time extent of star formation during the separation phase might not be found in a retrograde encounter.

Our results show that the observational detection or the absence of a sustained star formation activity in interacting galaxies, in particular over large volumes in tidal debris, would provide a strong hint on the nature of gravitation on galactic scales, as a complement to the study of rotation curves for isolated discs. Even when focussing on the merger remnant only, we show that the Milgromian paradigm favours a long and approximately continuous episode of star formation, starting at the first encounter and ending a few $100 \Myr$ after coalescence. Oppositely, the Newtonian formalism rather supports distinct bursts of star formation at high efficiency, associated with the close encounters \citep[see also][]{DiMatteo2007}. Providing that observational techniques would be able to tell apart these two star formation histories with sufficient confidence in the required age range \citep[$\approx 200 \Myr \mh 1 \Gyr$, see][on the difficulties of dating post-starburst episodes]{Lancon2001, Maraston2001, Maraston2004, Simones2014}, such interacting systems could be used as clear diagnostics to test different paradigms for gravitation on galactic scales.

%%%%%%%%%%%%%%%%%%%%%%%%%%%%%%%%%%%%%%%%%%%%%%%%%%
\section*{Acknowledgments}
We thank the referee for constructive comments which helped to improve the manuscript. This work was granted access to the PRACE Research Infrastructure resource \emph{Curie} hosted at the TGCC (France). FR acknowledges support from the European Research Council through grant ERC-StG-335936. BF acknowledges the financial support from the \emph{Programme Investissements d'Avenir} (PIA) of the \emph{Initiative d'Excellence} (IdEx) program from Universit\'e de Strasbourg.

%%%%%%%%%%%%%%%%%%%%%%%%%%%%%%%%%%%%%%%%%%%%%%%%%%%%%%%%%%%%%%%%%%%%%%%%%%%%%%%%%%%%%%%%%%%%%%%%%%%%%%%%%%%%%%%%%%%%%%%%%%%%%%%%%%%%%%%%%%%%%%%%%%%%%%%%%%%%%%%%%%%%%%%%%%%%%%%%%%%%%%

\bibliographystyle{mn2e}

\begin{thebibliography}{}

\end{thebibliography}


\begin{thebibliography}{}

\bibitem[\protect\citeauthoryear{{Angus}, {van der Heyden}, {Famaey},
  {Gentile}, {McGaugh} \& {de Blok}}{{Angus} et~al.}{2012}]{Angus2012}
{Angus} G.~W.,  {van der Heyden} K.~J.,  {Famaey} B.,  {Gentile} G.,  {McGaugh}
  S.~S.,    {de Blok} W.~J.~G.,  2012, \mnras, 421, 2598

\bibitem[\protect\citeauthoryear{{Armus}, {Heckman} \& {Miley}}{{Armus}
  et~al.}{1987}]{Armus1987}
{Armus} L.,  {Heckman} T.,    {Miley} G.,  1987, \aj, 94, 831

\bibitem[\protect\citeauthoryear{{Arp}}{{Arp}}{1966}]{Arp1966}
{Arp} H.,  1966, {Atlas of peculiar galaxies}

\bibitem[\protect\citeauthoryear{{Barnes}}{{Barnes}}{2004}]{Barnes2004}
{Barnes} J.~E.,  2004, \mnras, 350, 798

\bibitem[\protect\citeauthoryear{{Barnes} \& {Hernquist}}{{Barnes} \&
  {Hernquist}}{1992}]{Barnes1992}
{Barnes} J.~E.,  {Hernquist} L.,  1992, \araa, 30, 705

\bibitem[\protect\citeauthoryear{{Barnes} \& {Hernquist}}{{Barnes} \&
  {Hernquist}}{1991}]{Barnes1991}
{Barnes} J.~E.,  {Hernquist} L.~E.,  1991, \apjl, 370, L65

\bibitem[\protect\citeauthoryear{{Bate} \& {Bonnell}}{{Bate} \&
  {Bonnell}}{2005}]{Bate2005}
{Bate} M.~R.,  {Bonnell} I.~A.,  2005, \mnras, 356, 1201

\bibitem[\protect\citeauthoryear{{Bonnell}, {Dobbs} \& {Smith}}{{Bonnell}
  et~al.}{2013}]{Bonnell2013}
{Bonnell} I.~A.,  {Dobbs} C.~L.,    {Smith} R.~J.,  2013, \mnras, 430, 1790

\bibitem[\protect\citeauthoryear{{Boquien}, {Lisenfeld}, {Duc}, {Braine},
  {Bournaud}, {Brinks} \& {Charmandaris}}{{Boquien} et~al.}{2011}]{Boquien2011}
{Boquien} M.,  {Lisenfeld} U.,  {Duc} P.-A.,  {Braine} J.,  {Bournaud} F.,
  {Brinks} E.,    {Charmandaris} V.,  2011, \aap, 533, A19

\bibitem[\protect\citeauthoryear{{Bournaud}}{{Bournaud}}{2010}]{Bournaud2010a}
{Bournaud} F.,  2010, in {B.~Smith, J.~Higdon, S.~Higdon, \& N.~Bastian} ed.,
  Galaxy Wars: Stellar Populations and Star Formation in Interacting Galaxies
  Vol.~423 of Astronomical Society of the Pacific Conference Series, {Star
  Formation and Structure Formation in Galaxy Interactions and Mergers}.
pp 177--+

\bibitem[\protect\citeauthoryear{{Bournaud}, {Duc}, {Amram}, {Combes} \&
  {Gach}}{{Bournaud} et~al.}{2004}]{Bournaud2004}
{Bournaud} F.,  {Duc} P.-A.,  {Amram} P.,  {Combes} F.,    {Gach} J.-L.,  2004,
  \aap, 425, 813

\bibitem[\protect\citeauthoryear{{Bournaud}, {Duc} \& {Masset}}{{Bournaud}
  et~al.}{2003}]{Bournaud2003}
{Bournaud} F.,  {Duc} P.-A.,    {Masset} F.,  2003, \aap, 411, L469

\bibitem[\protect\citeauthoryear{{Boylan-Kolchin}, {Bullock} \&
  {Kaplinghat}}{{Boylan-Kolchin} et~al.}{2011}]{BoylanKolchin2011}
{Boylan-Kolchin} M.,  {Bullock} J.~S.,    {Kaplinghat} M.,  2011, \mnras, 415,
  L40

\bibitem[\protect\citeauthoryear{{Brada} \& {Milgrom}}{{Brada} \&
  {Milgrom}}{1999}]{Brada1999}
{Brada} R.,  {Milgrom} M.,  1999, \apj, 519, 590

\bibitem[\protect\citeauthoryear{{Candlish}, {Smith} \& {Fellhauer}}{{Candlish}
  et~al.}{2015}]{Candlish2015}
{Candlish} G.~N.,  {Smith} R.,    {Fellhauer} M.,  2015, \mnras, 446, 1060

\bibitem[\protect\citeauthoryear{{Chapon}}{{Chapon}}{2011}]{Chapon2011}
{Chapon} D.,  2011, {Simulations of binary galaxy mergers : the effect of gas
  physics on small scales}.
PhD Thesis, Universite Paris 7, 2011

\bibitem[\protect\citeauthoryear{{Chien} \& {Barnes}}{{Chien} \&
  {Barnes}}{2010}]{Chien2010}
{Chien} L.-H.,  {Barnes} J.~E.,  2010, \mnras, 407, 43

\bibitem[\protect\citeauthoryear{{Chien}, {Barnes}, {Kewley} \&
  {Chambers}}{{Chien} et~al.}{2007}]{Chien2007}
{Chien} L.-H.,  {Barnes} J.~E.,  {Kewley} L.~J.,    {Chambers} K.~C.,  2007,
  \apjl, 660, L105

\bibitem[\protect\citeauthoryear{{Chou}, {Bridge} \& {Abraham}}{{Chou}
  et~al.}{2012}]{Chou2012}
{Chou} R.~C.~Y.,  {Bridge} C.~R.,    {Abraham} R.~G.,  2012, \apj, 760, 113

\bibitem[\protect\citeauthoryear{{Combes} \& {Tiret}}{{Combes} \&
  {Tiret}}{2010}]{Combes2010}
{Combes} F.,  {Tiret} O.,  2010, in {Alimi} J.-M.,  {Fu{\"o}zfa} A.,  eds,
  American Institute of Physics Conference Series Vol.~1241 of American
  Institute of Physics Conference Series, {MOND and the Galaxies}.
pp 154--161

\bibitem[\protect\citeauthoryear{{Courty} \& {Alimi}}{{Courty} \&
  {Alimi}}{2004}]{Courty2004}
{Courty} S.,  {Alimi} J.~M.,  2004, \aap, 416, 875

\bibitem[\protect\citeauthoryear{{Daddi}, {Bournaud}, {Walter}, {Dannerbauer},
  {Carilli}, {Dickinson}, {Elbaz}, {Morrison} \& {et al.}}{{Daddi}
  et~al.}{2010}]{Daddi2010a}
{Daddi} E.,  {Bournaud} F.,  {Walter} F.,  {Dannerbauer} H.,  {Carilli} C.~L.,
  {Dickinson} M.,  {Elbaz} D.,  {Morrison} G.~E.,    {et al.} 2010, \apj, 713,
  686

\bibitem[\protect\citeauthoryear{{Daddi}, {Elbaz}, {Walter}, {Bournaud},
  {Salmi}, {Carilli}, {Dannerbauer}, {Dickinson} \& {et al.}}{{Daddi}
  et~al.}{2010}]{Daddi2010b}
{Daddi} E.,  {Elbaz} D.,  {Walter} F.,  {Bournaud} F.,  {Salmi} F.,  {Carilli}
  C.,  {Dannerbauer} H.,  {Dickinson} M.,    {et al.} 2010, \apjl, 714, L118

\bibitem[\protect\citeauthoryear{{de Blok}}{{de Blok}}{2010}]{deBlok2010}
{de Blok} W.~J.~G.,  2010, Advances in Astronomy, 2010, 789293

\bibitem[\protect\citeauthoryear{{Dehnen}, {McLaughlin} \& {Sachania}}{{Dehnen}
  et~al.}{2006}]{Dehnen2006}
{Dehnen} W.,  {McLaughlin} D.~E.,    {Sachania} J.,  2006, \mnras, 369, 1688

\bibitem[\protect\citeauthoryear{{Di Matteo}, {Combes}, {Melchior} \&
  {Semelin}}{{Di Matteo} et~al.}{2007}]{DiMatteo2007}
{Di Matteo} P.,  {Combes} F.,  {Melchior} A.,    {Semelin} B.,  2007, \aap,
  468, 61

\bibitem[\protect\citeauthoryear{{Dubinski}, {Mihos} \& {Hernquist}}{{Dubinski}
  et~al.}{1996}]{Dubinski1996}
{Dubinski} J.,  {Mihos} J.~C.,    {Hernquist} L.,  1996, \apj, 462, 576

\bibitem[\protect\citeauthoryear{{Dubois} \& {Teyssier}}{{Dubois} \&
  {Teyssier}}{2008}]{Dubois2008}
{Dubois} Y.,  {Teyssier} R.,  2008, \aap, 477, 79

\bibitem[\protect\citeauthoryear{{Duc} \& {Renaud}}{{Duc} \&
  {Renaud}}{2013}]{Duc2013}
{Duc} P.-A.,  {Renaud} F.,  2013, in {Souchay} J.,  {Mathis} S.,   {Tokieda}
  T.,  eds, Lecture Notes in Physics, Berlin Springer Verlag Vol.~861, {Tides
  in Colliding Galaxies}.
p.~327

\bibitem[\protect\citeauthoryear{{Ellison}, {Mendel}, {Patton} \&
  {Scudder}}{{Ellison} et~al.}{2013}]{Ellison2013}
{Ellison} S.~L.,  {Mendel} J.~T.,  {Patton} D.~R.,    {Scudder} J.~M.,  2013,
  \mnras, 435, 3627

\bibitem[\protect\citeauthoryear{{Famaey} \& {Binney}}{{Famaey} \&
  {Binney}}{2005}]{Famaey2005}
{Famaey} B.,  {Binney} J.,  2005, \mnras, 363, 603

\bibitem[\protect\citeauthoryear{{Famaey} \& {McGaugh}}{{Famaey} \&
  {McGaugh}}{2012}]{Famaey2012}
{Famaey} B.,  {McGaugh} S.~S.,  2012, Living Reviews in Relativity, 15, 10

\bibitem[\protect\citeauthoryear{{Flores}, {Hammer}, {Fouquet}, {Puech},
  {Kroupa}, {Yang} \& {Pawlowski}}{{Flores} et~al.}{2016}]{Flores2016}
{Flores} H.,  {Hammer} F.,  {Fouquet} S.,  {Puech} M.,  {Kroupa} P.,  {Yang}
  Y.,    {Pawlowski} M.,  2016, \mnras, 457, L14

\bibitem[\protect\citeauthoryear{{Gentile}, {Famaey}, {Combes}, {Kroupa},
  {Zhao} \& {Tiret}}{{Gentile} et~al.}{2007}]{Gentile2007}
{Gentile} G.,  {Famaey} B.,  {Combes} F.,  {Kroupa} P.,  {Zhao} H.~S.,
  {Tiret} O.,  2007, \aap, 472, L25

\bibitem[\protect\citeauthoryear{{Genzel}, {Tacconi}, {Gracia-Carpio},
  {Sternberg}, {Cooper}, {Shapiro}, {Bolatto} \& {et al.}}{{Genzel}
  et~al.}{2010}]{Genzel2010}
{Genzel} R.,  {Tacconi} L.~J.,  {Gracia-Carpio} J.,  {Sternberg} A.,  {Cooper}
  M.~C.,  {Shapiro} K.,  {Bolatto} A.,    {et al.} 2010, \mnras, 407, 2091

\bibitem[\protect\citeauthoryear{{Hees}, {Famaey}, {Angus} \& {Gentile}}{{Hees}
  et~al.}{2016}]{Hees2016}
{Hees} A.,  {Famaey} B.,  {Angus} G.~W.,    {Gentile} G.,  2016, \mnras, 455,
  449

\bibitem[\protect\citeauthoryear{{Hennebelle} \& {Chabrier}}{{Hennebelle} \&
  {Chabrier}}{2008}]{Hennebelle2008}
{Hennebelle} P.,  {Chabrier} G.,  2008, \apj, 684, 395

\bibitem[\protect\citeauthoryear{{Hibbard} \& {Mihos}}{{Hibbard} \&
  {Mihos}}{1995}]{Hibbard1995}
{Hibbard} J.~E.,  {Mihos} J.~C.,  1995, \aj, 110, 140

\bibitem[\protect\citeauthoryear{{Hibbard}, {van der Hulst}, {Barnes} \&
  {Rich}}{{Hibbard} et~al.}{2001}]{Hibbard2001}
{Hibbard} J.~E.,  {van der Hulst} J.~M.,  {Barnes} J.~E.,    {Rich} R.~M.,
  2001, \aj, 122, 2969

\bibitem[\protect\citeauthoryear{{Hopkins}}{{Hopkins}}{2013}]{Hopkins2013}
{Hopkins} P.~F.,  2013, \mnras, 430, 1880

\bibitem[\protect\citeauthoryear{{Hopkins}, {Cox}, {Younger} \&
  {Hernquist}}{{Hopkins} et~al.}{2009}]{Hopkins2009}
{Hopkins} P.~F.,  {Cox} T.~J.,  {Younger} J.~D.,    {Hernquist} L.,  2009,
  \apj, 691, 1168

\bibitem[\protect\citeauthoryear{{Houck}, {Schneider}, {Danielson},
  {Neugebauer}, {Soifer}, {Beichman} \& {Lonsdale}}{{Houck}
  et~al.}{1985}]{Houck1985}
{Houck} J.~R.,  {Schneider} D.~P.,  {Danielson} G.~E.,  {Neugebauer} G.,
  {Soifer} B.~T.,  {Beichman} C.~A.,    {Lonsdale} C.~J.,  1985, \apjl, 290, L5

\bibitem[\protect\citeauthoryear{{Ibata}, {Ibata}, {Famaey} \& {Lewis}}{{Ibata}
  et~al.}{2014}]{Ibata2014}
{Ibata} N.~G.,  {Ibata} R.~A.,  {Famaey} B.,    {Lewis} G.~F.,  2014, \nat,
  511, 563

\bibitem[\protect\citeauthoryear{{Ibata}, {Lewis}, {Conn}, {Irwin},
  {McConnachie}, {Chapman}, {Collins}, {Fardal} \& {et al.}}{{Ibata}
  et~al.}{2013}]{Ibata2013}
{Ibata} R.~A.,  {Lewis} G.~F.,  {Conn} A.~R.,  {Irwin} M.~J.,  {McConnachie}
  A.~W.,  {Chapman} S.~C.,  {Collins} M.~L.,  {Fardal} M.,    {et al.} 2013,
  \nat, 493, 62

\bibitem[\protect\citeauthoryear{{Jog}}{{Jog}}{2013}]{Jog2013}
{Jog} C.~J.,  2013, \mnras, 434, L56

\bibitem[\protect\citeauthoryear{{Jog}}{{Jog}}{2014}]{Jog2014}
{Jog} C.~J.,  2014, \aj, 147, 132

\bibitem[\protect\citeauthoryear{{Jog} \& {Solomon}}{{Jog} \&
  {Solomon}}{1992}]{Jog1992}
{Jog} C.~J.,  {Solomon} P.~M.,  1992, \apj, 387, 152

\bibitem[\protect\citeauthoryear{{Kennicutt}}{{Kennicutt}}{1998}]{Kennicutt1998b}
{Kennicutt} R.~C.,  1998, \apj, 498, 541

\bibitem[\protect\citeauthoryear{{Kroupa}}{{Kroupa}}{2015}]{Kroupa2015}
{Kroupa} P.,  2015, Canadian Journal of Physics, 93, 169

\bibitem[\protect\citeauthoryear{{Kroupa}, {Famaey}, {de Boer},
  {Dabringhausen}, {Pawlowski}, {Boily}, {Jerjen}, {Forbes} \& {et
  al.}}{{Kroupa} et~al.}{2010}]{Kroupa2010}
{Kroupa} P.,  {Famaey} B.,  {de Boer} K.~S.,  {Dabringhausen} J.,  {Pawlowski}
  M.~S.,  {Boily} C.~M.,  {Jerjen} H.,  {Forbes} D.,    {et al.} 2010, \aap,
  523, A32

\bibitem[\protect\citeauthoryear{{Kroupa}, {Theis} \& {Boily}}{{Kroupa}
  et~al.}{2005}]{Kroupa2005b}
{Kroupa} P.,  {Theis} C.,    {Boily} C.~M.,  2005, \aap, 431, 517

\bibitem[\protect\citeauthoryear{{Kuijken} \& {Dubinski}}{{Kuijken} \&
  {Dubinski}}{1995}]{Kuijken1995}
{Kuijken} K.,  {Dubinski} J.,  1995, \mnras, 277, 1341

\bibitem[\protect\citeauthoryear{{Lan{\c c}on}}{{Lan{\c
  c}on}}{2001}]{Lancon2001}
{Lan{\c c}on} A.,  2001, in {Tacconi} L.,  {Lutz} D.,  eds, Starburst Galaxies:
  Near and Far {Can We Date Starbursts?}.
p.~231

\bibitem[\protect\citeauthoryear{{Lelli}, {Duc}, {Brinks}, {Bournaud},
  {McGaugh}, {Lisenfeld}, {Weilbacher}, {Boquien} \& {et al.}}{{Lelli}
  et~al.}{2015}]{Lelli2015}
{Lelli} F.,  {Duc} P.-A.,  {Brinks} E.,  {Bournaud} F.,  {McGaugh} S.~S.,
  {Lisenfeld} U.,  {Weilbacher} P.~M.,  {Boquien} M.,    {et al.} 2015, \aap,
  584, A113

\bibitem[\protect\citeauthoryear{{Lelli}, {Fraternali} \& {Verheijen}}{{Lelli}
  et~al.}{2013}]{Lelli2013}
{Lelli} F.,  {Fraternali} F.,    {Verheijen} M.,  2013, \mnras, 433, L30

\bibitem[\protect\citeauthoryear{{Lelli}, {McGaugh} \& {Schombert}}{{Lelli}
  et~al.}{2016a}]{Lelli2016a}
{Lelli} F.,  {McGaugh} S.~S.,    {Schombert} J.~M.,  2016a, \apjl, 816, L14

\bibitem[\protect\citeauthoryear{{Lelli}, {McGaugh}, {Schombert} \&
  {Pawlowski}}{{Lelli} et~al.}{2016b}]{Lelli2016b}
{Lelli} F.,  {McGaugh} S.~S.,  {Schombert} J.~M.,    {Pawlowski} M.~S.,  2016b,
  \apjl, 827, L19

\bibitem[\protect\citeauthoryear{{Llinares}, {Knebe} \& {Zhao}}{{Llinares}
  et~al.}{2008}]{Llinares2008}
{Llinares} C.,  {Knebe} A.,    {Zhao} H.,  2008, \mnras, 391, 1778

\bibitem[\protect\citeauthoryear{{Loren}}{{Loren}}{1976}]{Loren1976}
{Loren} R.~B.,  1976, \apj, 209, 466

\bibitem[\protect\citeauthoryear{{L{\"u}ghausen}, {Famaey} \&
  {Kroupa}}{{L{\"u}ghausen} et~al.}{2015}]{Lughausen2015}
{L{\"u}ghausen} F.,  {Famaey} B.,    {Kroupa} P.,  2015, Canadian Journal of
  Physics, 93, 232

\bibitem[\protect\citeauthoryear{{Mac Low} \& {Klessen}}{{Mac Low} \&
  {Klessen}}{2004}]{MacLow2004}
{Mac Low} M.-M.,  {Klessen} R.~S.,  2004, Reviews of Modern Physics, 76, 125

\bibitem[\protect\citeauthoryear{{Maraston}, {Bastian}, {Saglia},
  {Kissler-Patig}, {Schweizer} \& {Goudfrooij}}{{Maraston}
  et~al.}{2004}]{Maraston2004}
{Maraston} C.,  {Bastian} N.,  {Saglia} R.~P.,  {Kissler-Patig} M.,
  {Schweizer} F.,    {Goudfrooij} P.,  2004, \aap, 416, 467

\bibitem[\protect\citeauthoryear{{Maraston}, {Kissler-Patig}, {Brodie},
  {Barmby} \& {Huchra}}{{Maraston} et~al.}{2001}]{Maraston2001}
{Maraston} C.,  {Kissler-Patig} M.,  {Brodie} J.~P.,  {Barmby} P.,    {Huchra}
  J.~P.,  2001, \aap, 370, 176

\bibitem[\protect\citeauthoryear{{McGaugh}, {Schombert}, {Bothun} \& {de
  Blok}}{{McGaugh} et~al.}{2000}]{McGaugh2000}
{McGaugh} S.~S.,  {Schombert} J.~M.,  {Bothun} G.~D.,    {de Blok} W.~J.~G.,
  2000, \apjl, 533, L99

\bibitem[\protect\citeauthoryear{{Metz}, {Kroupa} \& {Jerjen}}{{Metz}
  et~al.}{2007}]{Metz2007}
{Metz} M.,  {Kroupa} P.,    {Jerjen} H.,  2007, \mnras, 374, 1125

\bibitem[\protect\citeauthoryear{{Metz}, {Kroupa} \& {Libeskind}}{{Metz}
  et~al.}{2008}]{Metz2008}
{Metz} M.,  {Kroupa} P.,    {Libeskind} N.~I.,  2008, \apj, 680, 287

\bibitem[\protect\citeauthoryear{{Mihos}, {Dubinski} \& {Hernquist}}{{Mihos}
  et~al.}{1998}]{Mihos1998}
{Mihos} J.~C.,  {Dubinski} J.,    {Hernquist} L.,  1998, \apj, 494, 183

\bibitem[\protect\citeauthoryear{{Milgrom}}{{Milgrom}}{1983}]{Milgrom1983}
{Milgrom} M.,  1983, \apj, 270, 365

\bibitem[\protect\citeauthoryear{{Milgrom}}{{Milgrom}}{2010}]{Milgrom2010}
{Milgrom} M.,  2010, \mnras, 403, 886

\bibitem[\protect\citeauthoryear{{Milgrom}}{{Milgrom}}{2016}]{Milgrom2016}
{Milgrom} M.,  2016, ArXiv e-prints

\bibitem[\protect\citeauthoryear{{Mirabel}, {Dottori} \& {Lutz}}{{Mirabel}
  et~al.}{1992}]{Mirabel1992}
{Mirabel} I.~F.,  {Dottori} H.,    {Lutz} D.,  1992, \aap, 256, L19

\bibitem[\protect\citeauthoryear{{Mondal} \& {Chakraborty}}{{Mondal} \&
  {Chakraborty}}{2015}]{Mondal2015}
{Mondal} S.,  {Chakraborty} S.,  2015, \mnras, 450, 1874

\bibitem[\protect\citeauthoryear{{Moreno}, {Bluck}, {Ellison}, {Patton},
  {Torrey} \& {Moster}}{{Moreno} et~al.}{2013}]{Moreno2013}
{Moreno} J.,  {Bluck} A.~F.~L.,  {Ellison} S.~L.,  {Patton} D.~R.,  {Torrey}
  P.,    {Moster} B.~P.,  2013, \mnras, 436, 1765

\bibitem[\protect\citeauthoryear{{Navarro}, {Frenk} \& {White}}{{Navarro}
  et~al.}{1997}]{Navarro1997}
{Navarro} J.~F.,  {Frenk} C.~S.,    {White} S.~D.~M.,  1997, \apj, 490, 493

\bibitem[\protect\citeauthoryear{{Nipoti}, {Ciotti}, {Binney} \&
  {Londrillo}}{{Nipoti} et~al.}{2008}]{Nipoti2008}
{Nipoti} C.,  {Ciotti} L.,  {Binney} J.,    {Londrillo} P.,  2008, \mnras, 386,
  2194

\bibitem[\protect\citeauthoryear{{Nipoti}, {Londrillo} \& {Ciotti}}{{Nipoti}
  et~al.}{2007}]{Nipoti2007}
{Nipoti} C.,  {Londrillo} P.,    {Ciotti} L.,  2007, \apj, 660, 256

\bibitem[\protect\citeauthoryear{{Oman}, {Navarro}, {Fattahi}, {Frenk},
  {Sawala}, {White}, {Bower}, {Crain} \& {et al.}}{{Oman}
  et~al.}{2015}]{Oman2015}
{Oman} K.~A.,  {Navarro} J.~F.,  {Fattahi} A.,  {Frenk} C.~S.,  {Sawala} T.,
  {White} S.~D.~M.,  {Bower} R.,  {Crain} R.~A.,    {et al.} 2015, \mnras, 452,
  3650

\bibitem[\protect\citeauthoryear{{Papastergis}, {Adams} \& {van der
  Hulst}}{{Papastergis} et~al.}{2016}]{Papastergis2016}
{Papastergis} E.,  {Adams} E.~A.~K.,    {van der Hulst} J.~M.,  2016, ArXiv
  e-prints

\bibitem[\protect\citeauthoryear{{Papastergis}, {Giovanelli}, {Haynes} \&
  {Shankar}}{{Papastergis} et~al.}{2015}]{Papastergis2015}
{Papastergis} E.,  {Giovanelli} R.,  {Haynes} M.~P.,    {Shankar} F.,  2015,
  \aap, 574, A113

\bibitem[\protect\citeauthoryear{{Pawlowski}, {Famaey}, {Merritt} \&
  {Kroupa}}{{Pawlowski} et~al.}{2015}]{Pawlowski2015}
{Pawlowski} M.~S.,  {Famaey} B.,  {Merritt} D.,    {Kroupa} P.,  2015, \apj,
  815, 19

\bibitem[\protect\citeauthoryear{{Pawlowski}, {Pflamm-Altenburg} \&
  {Kroupa}}{{Pawlowski} et~al.}{2012}]{Pawlowski2012}
{Pawlowski} M.~S.,  {Pflamm-Altenburg} J.,    {Kroupa} P.,  2012, \mnras, 423,
  1109

\bibitem[\protect\citeauthoryear{{Privon}, {Barnes}, {Evans}, {Hibbard}, {Yun},
  {Mazzarella}, {Armus} \& {Surace}}{{Privon} et~al.}{2013}]{Privon2013}
{Privon} G.~C.,  {Barnes} J.~E.,  {Evans} A.~S.,  {Hibbard} J.~E.,  {Yun}
  M.~S.,  {Mazzarella} J.~M.,  {Armus} L.,    {Surace} J.,  2013, \apj, 771,
  120

\bibitem[\protect\citeauthoryear{{Quinn}, {Hernquist} \& {Fullagar}}{{Quinn}
  et~al.}{1993}]{Quinn1993}
{Quinn} P.~J.,  {Hernquist} L.,    {Fullagar} D.~P.,  1993, \apj, 403, 74

\bibitem[\protect\citeauthoryear{{Read}, {Iorio}, {Agertz} \&
  {Fraternali}}{{Read} et~al.}{2016}]{Read2016b}
{Read} J.~I.,  {Iorio} G.,  {Agertz} O.,    {Fraternali} F.,  2016, ArXiv
  e-prints

\bibitem[\protect\citeauthoryear{{Renaud}, {Boily}, {Fleck}, {Naab} \&
  {Theis}}{{Renaud} et~al.}{2008}]{Renaud2008}
{Renaud} F.,  {Boily} C.~M.,  {Fleck} J.-J.,  {Naab} T.,    {Theis} C.,  2008,
  \mnras, 391, L98

\bibitem[\protect\citeauthoryear{{Renaud}, {Boily}, {Naab} \& {Theis}}{{Renaud}
  et~al.}{2009}]{Renaud2009}
{Renaud} F.,  {Boily} C.~M.,  {Naab} T.,    {Theis} C.,  2009, \apj, 706, 67

\bibitem[\protect\citeauthoryear{{Renaud}, {Bournaud} \& {Duc}}{{Renaud}
  et~al.}{2015a}]{Renaud2015a}
{Renaud} F.,  {Bournaud} F.,    {Duc} P.-A.,  2015a, \mnras, 446, 2038

\bibitem[\protect\citeauthoryear{{Renaud}, {Bournaud}, {Emsellem}, {Agertz},
  {Athanassoula}, {Combes}, {Elmegreen}, {Kraljic}, {Motte} \&
  {Teyssier}}{{Renaud} et~al.}{2015b}]{Renaud2015d}
{Renaud} F.,  {Bournaud} F.,  {Emsellem} E.,  {Agertz} O.,  {Athanassoula} E.,
  {Combes} F.,  {Elmegreen} B.,  {Kraljic} K.,  {Motte} F.,    {Teyssier} R.,
  2015b, \mnras, 454, 3299

\bibitem[\protect\citeauthoryear{{Renaud}, {Bournaud}, {Emsellem}, {Elmegreen},
  {Teyssier}, {Alves}, {Chapon}, {Combes} \& {et al.}}{{Renaud}
  et~al.}{2013}]{Renaud2013b}
{Renaud} F.,  {Bournaud} F.,  {Emsellem} E.,  {Elmegreen} B.,  {Teyssier} R.,
  {Alves} J.,  {Chapon} D.,  {Combes} F.,    {et al.} 2013, \mnras, 436, 1836

\bibitem[\protect\citeauthoryear{{Renaud}, {Bournaud}, {Kraljic} \&
  {Duc}}{{Renaud} et~al.}{2014}]{Renaud2014b}
{Renaud} F.,  {Bournaud} F.,  {Kraljic} K.,    {Duc} P.-A.,  2014, \mnras, 442,
  L33

\bibitem[\protect\citeauthoryear{{Saintonge}, {Tacconi}, {Fabello}, {Wang},
  {Catinella}, {Genzel}, {Graci{\'a}-Carpio}, {Kramer} \& {et al.}}{{Saintonge}
  et~al.}{2012}]{Saintonge2012}
{Saintonge} A.,  {Tacconi} L.~J.,  {Fabello} S.,  {Wang} J.,  {Catinella} B.,
  {Genzel} R.,  {Graci{\'a}-Carpio} J.,  {Kramer} C.,    {et al.} 2012, \apj,
  758, 73

\bibitem[\protect\citeauthoryear{{Sanders} \& {Mirabel}}{{Sanders} \&
  {Mirabel}}{1996}]{Sanders1996}
{Sanders} D.~B.,  {Mirabel} I.~F.,  1996, \araa, 34, 749

\bibitem[\protect\citeauthoryear{{Schombert}, {Wallin} \&
  {Struck-Marcell}}{{Schombert} et~al.}{1990}]{Schombert1990}
{Schombert} J.~M.,  {Wallin} J.~F.,    {Struck-Marcell} C.,  1990, \aj, 99, 497

\bibitem[\protect\citeauthoryear{{Schreiber}, {Pannella}, {Elbaz},
  {B{\'e}thermin}, {Inami}, {Dickinson}, {Magnelli}, {Wang} \& {et
  al.}}{{Schreiber} et~al.}{2015}]{Schreiber2015}
{Schreiber} C.,  {Pannella} M.,  {Elbaz} D.,  {B{\'e}thermin} M.,  {Inami} H.,
  {Dickinson} M.,  {Magnelli} B.,  {Wang} T.,    {et al.} 2015, \aap, 575, A74

\bibitem[\protect\citeauthoryear{{Scudder}, {Ellison}, {Torrey}, {Patton} \&
  {Mendel}}{{Scudder} et~al.}{2012}]{Scudder2012}
{Scudder} J.~M.,  {Ellison} S.~L.,  {Torrey} P.,  {Patton} D.~R.,    {Mendel}
  J.~T.,  2012, \mnras, 426, 549

\bibitem[\protect\citeauthoryear{{Simones}, {Weisz}, {Skillman}, {Bell},
  {Bianchi}, {Dalcanton}, {Dolphin}, {Johnson} \& {Williams}}{{Simones}
  et~al.}{2014}]{Simones2014}
{Simones} J.~E.,  {Weisz} D.~R.,  {Skillman} E.~D.,  {Bell} E.~F.,  {Bianchi}
  L.,  {Dalcanton} J.~J.,  {Dolphin} A.~E.,  {Johnson} B.~D.,    {Williams}
  B.~F.,  2014, \apj, 788, 12

\bibitem[\protect\citeauthoryear{{Smith}, {Carver}, {Pfeiffer}, {Perkins},
  {Barkanic}, {Fritts}, {Southerland}, {Manchikalapudi}, {Baker}, {Luckey},
  {Franklin}, {Moffett} \& {Struck}}{{Smith} et~al.}{2010}]{Smith2010}
{Smith} B.~J.,  {Carver} D.~C.,  {Pfeiffer} P.,  {Perkins} S.,  {Barkanic} J.,
  {Fritts} S.,  {Southerland} D.,  {Manchikalapudi} D.,  {Baker} M.,  {Luckey}
  J.,  {Franklin} C.,  {Moffett} A.,    {Struck} C.,  2010, in {B.~Smith,
  J.~Higdon, S.~Higdon, \& N.~Bastian} ed., Galaxy Wars: Stellar Populations
  and Star Formation in Interacting Galaxies Vol.~423 of Astronomical Society
  of the Pacific Conference Series, {The Automatic Galaxy Collision Software}.
pp 227--+

\bibitem[\protect\citeauthoryear{{Springel} \& {White}}{{Springel} \&
  {White}}{1999}]{Springel1999}
{Springel} V.,  {White} S.~D.~M.,  1999, \mnras, 307, 162

\bibitem[\protect\citeauthoryear{{Struck}}{{Struck}}{1999}]{Struck1999}
{Struck} C.,  1999, \physrep, 321, 1

\bibitem[\protect\citeauthoryear{{Tan}}{{Tan}}{2000}]{Tan2000}
{Tan} J.~C.,  2000, \apj, 536, 173

\bibitem[\protect\citeauthoryear{{Teuben}}{{Teuben}}{1995}]{Teuben1995}
{Teuben} P.,  1995, in {Shaw} R.~A.,  {Payne} H.~E.,   {Hayes} J.~J.~E.,  eds,
  Astronomical Data Analysis Software and Systems IV Vol.~77 of Astronomical
  Society of the Pacific Conference Series, {The Stellar Dynamics Toolbox
  NEMO}.
p.~398

\bibitem[\protect\citeauthoryear{{Teyssier}}{{Teyssier}}{2002}]{Teyssier2002}
{Teyssier} R.,  2002, \aap, 385, 337

\bibitem[\protect\citeauthoryear{{Teyssier}, {Chapon} \& {Bournaud}}{{Teyssier}
  et~al.}{2010}]{Teyssier2010}
{Teyssier} R.,  {Chapon} D.,    {Bournaud} F.,  2010, \apjl, 720, L149

\bibitem[\protect\citeauthoryear{{Teyssier}, {Pontzen}, {Dubois} \&
  {Read}}{{Teyssier} et~al.}{2013}]{Teyssier2013}
{Teyssier} R.,  {Pontzen} A.,  {Dubois} Y.,    {Read} J.~I.,  2013, \mnras,
  429, 3068

\bibitem[\protect\citeauthoryear{{Tiret} \& {Combes}}{{Tiret} \&
  {Combes}}{2007}]{Tiret2007}
{Tiret} O.,  {Combes} F.,  2007, \aap, 464, 517

\bibitem[\protect\citeauthoryear{{Tiret} \& {Combes}}{{Tiret} \&
  {Combes}}{2008a}]{Tiret2008a}
{Tiret} O.,  {Combes} F.,  2008a, \aap, 483, 719

\bibitem[\protect\citeauthoryear{{Tiret} \& {Combes}}{{Tiret} \&
  {Combes}}{2008b}]{Tiret2008b}
{Tiret} O.,  {Combes} F.,  2008b, in {Funes} J.~G.,  {Corsini} E.~M.,  eds,
  Formation and Evolution of Galaxy Disks Vol.~396 of Astronomical Society of
  the Pacific Conference Series, {Interacting Galaxies with Modified Newtonian
  Dynamics}.
p.~259

\bibitem[\protect\citeauthoryear{{Toomre} \& {Toomre}}{{Toomre} \&
  {Toomre}}{1972}]{Toomre1972}
{Toomre} A.,  {Toomre} J.,  1972, \apj, 178, 623

\bibitem[\protect\citeauthoryear{{Wetzstein}, {Naab} \& {Burkert}}{{Wetzstein}
  et~al.}{2007}]{Wetzstein2007}
{Wetzstein} M.,  {Naab} T.,    {Burkert} A.,  2007, \mnras, 375, 805

\end{thebibliography}

%%%%%%%%%%%%%%%%%%%%%%%%%%%%%%%%%%%%%%%%%%%%%%%%%%
\appendix

%%%%%%%%%%%%%%%%%%%%%%%%%%%%%%%%%%%%%%%%%%%%%%%%%%%%%%%%%%%%%%%%%%%%%%%%%%%%%%%%
\section{Adaptive refinement}
\label{sec:refinement}

As mentioned in \sect{pb}, since \emph{one} of the refinement criteria used by \ramses and \por is based on the number of particles per cell, the absence of DM particles in the MOND simulations introduces differences in the effective resolution of our Newtonian and Milgromian runs.

\begin{figure}
\includegraphics{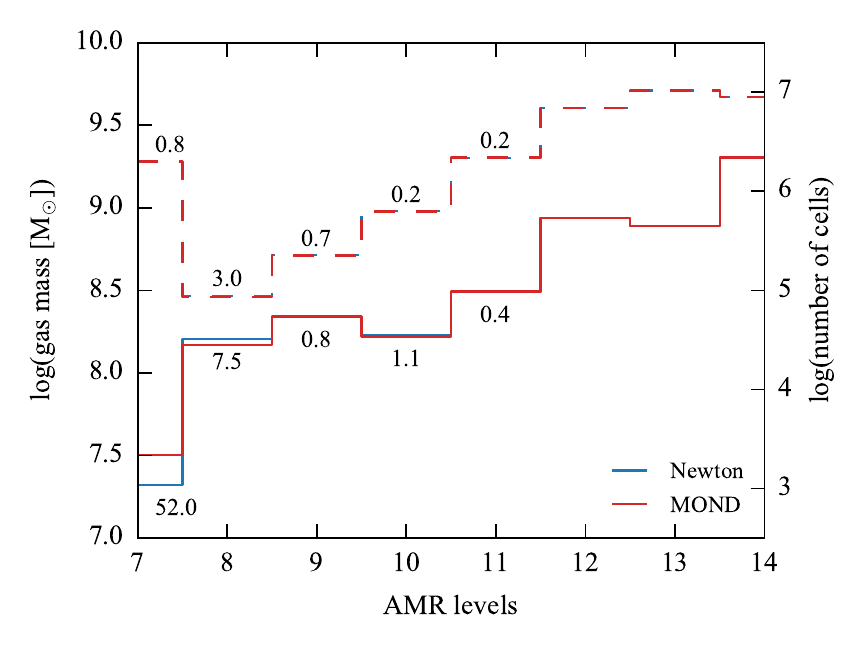}
\caption{Gas mass (solid lines, left axis) and number of cells (dashed lines, right axis) per refinement levels in simulations of an isolated galaxy (see text for details). The numbers indicate the percentage relative difference, i.e. $100\times$(1-MOND/Newton). For the finest levels, the relative difference (not shown) is smaller than $0.1 \%$. Since the simulation volume for these tests span $(100 \kpc)^3$, the size of the AMR cells is $100 / 2^\mathrm{level} \kpc$, i.e. from $781 \pc$ at level 7, to $6 \pc$ at level 14. For all the simulations presented here, the minimum refinement level is 7, meaning that the entire simulation volume is at least resolved at this level.}
\label{fig:levels}
\end{figure}

To quantify this effect as accurately as possible, we start a simulation of an isolated galaxy (as described in \sect{ic}) in the Newtonian framework, and evolve it for $100 \Myr$. At this point, we duplicate the simulation and keep one copy running. In the other one, we remove the DM particles and switch on the MOND Poisson solver. \fig{levels} shows the occupation of the refinement levels (only for the gas denser than $10^{-3} \cc$) in both runs, after another $50 \Myr$ of evolution.

The differences mainly arise in the coarsest levels (7 and 8), i.e. the diffuse interstellar and intergalactic media. This corresponds to volumes in the outermost regions of the galaxy, i.e. where the presence or absence of DM is the most relevant and dominates the refinement strategy. For finer levels, the gaseous and stellar components dominate the local density and thus the refinement is mostly governed by criteria mildly dependent on the presence of DM (the local Jeans length and the gas mass). Therefore, the differences between the two runs remain small.

%%%%%%%%%%%%%%%%%%%%%%%%%%%%%%%%%%%%%%%%%%%%%%%%%%%%%%%%%%%%%%%%%%%%%%%%%%%%%%%%
\section{Isolated galaxies}
\label{sec:isolated}

\begin{figure*}
\includegraphics{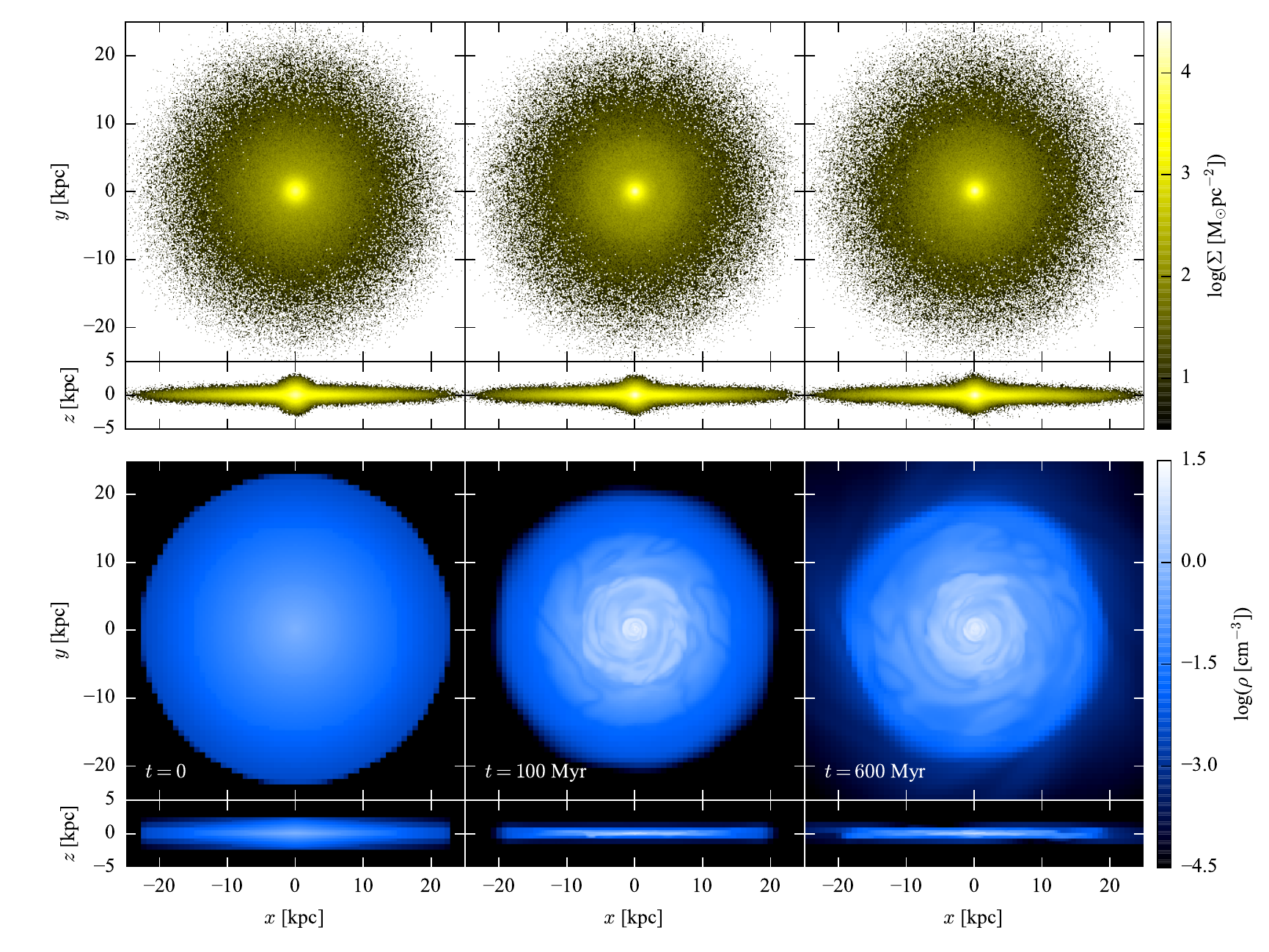}
\caption{Face-on and edge-on surface density map of the stellar component (top) and gas volume density averaged along the line of sight (bottom) for the Newtonian model run in isolation. The left column represents the initial conditions. The central and right columns are $100 \Myr$ and $600 \Myr$ later, showing the virialisation from the initial conditions, and the long term evolution, respectively.}
\label{fig:isomorpho_newton}
\end{figure*}

\begin{figure*}
\includegraphics{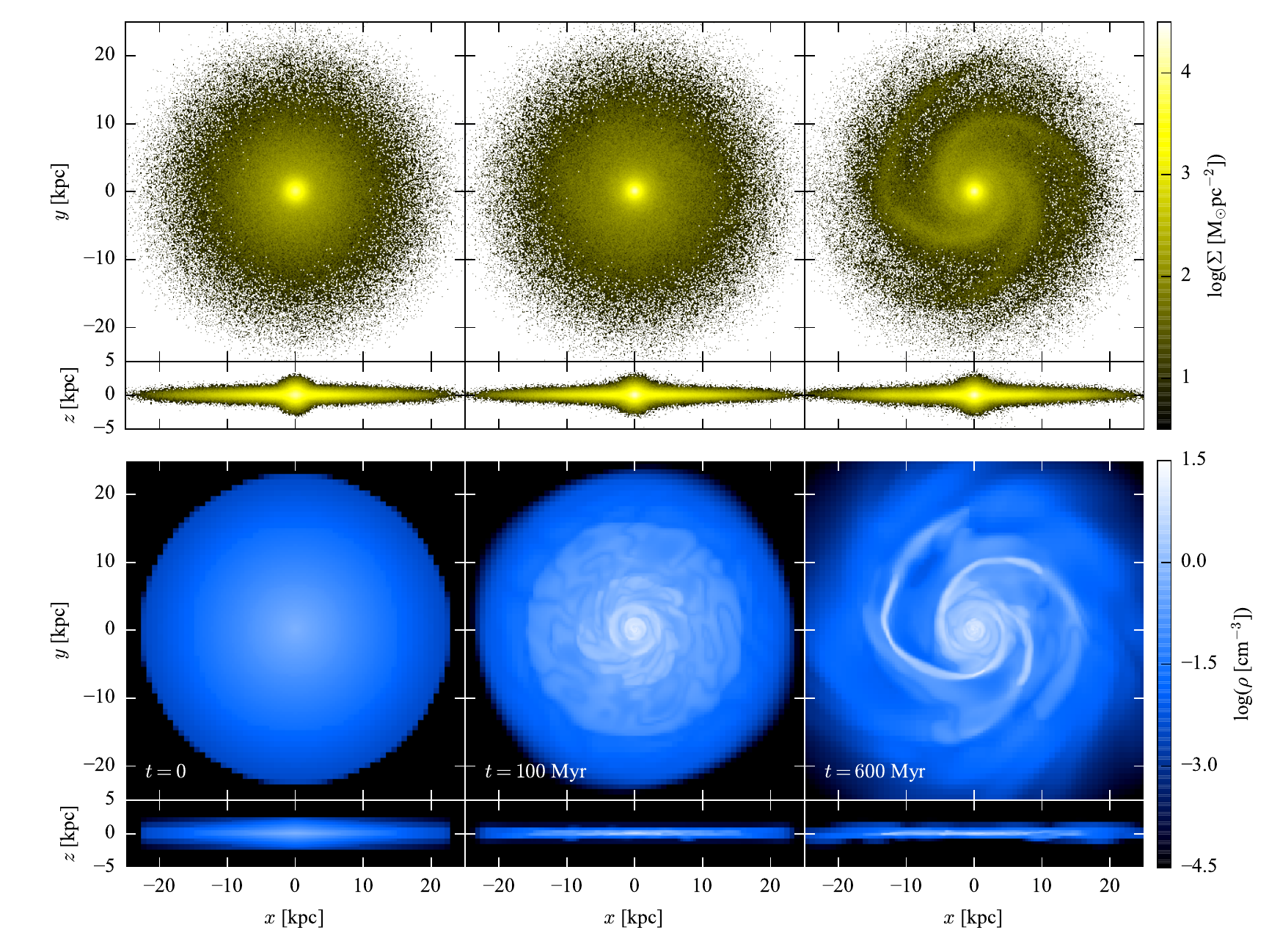}
\caption{Same as \fig{isomorpho_newton}, but for the Milgromian model. By construction, the initial conditions (left column) are strictly identical to the Newtonian case.}
\label{fig:isomorpho_mond}
\end{figure*}

\begin{figure}
\includegraphics{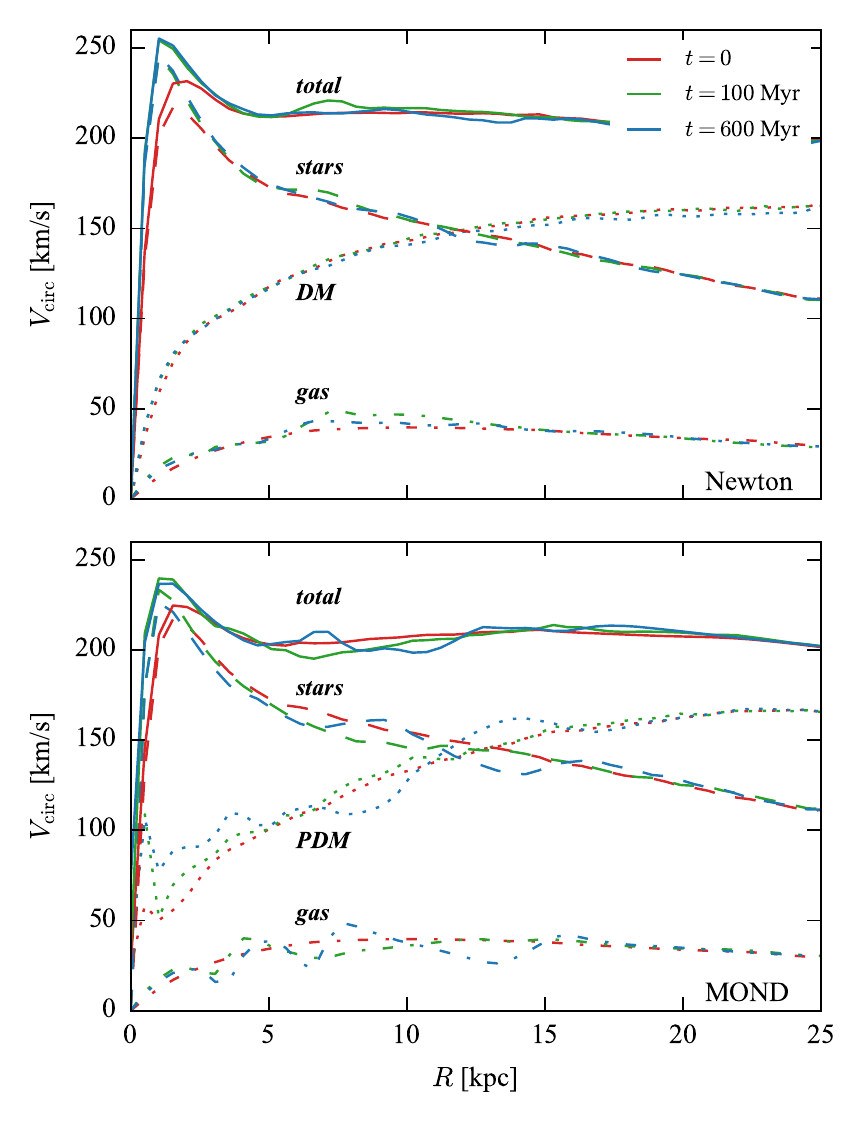}
\caption{Rotation curves (solid lines) of our Newtonian (top) and Milgromian (bottom) models at the three instants shown in \fig[s]{isomorpho_newton} and \ref{fig:isomorpho_mond}. Dashed and dash-dotted lines represent the rotation curves for the stellar and gaseous components respectively. The dotted lines correspond to the dark matter (Newton) and the phantom dark matter (MOND). The latter is computed by subtracting the baryonic contributions to the total circular velocity (assuming Newtonian physics), itself computed from the mid-plane total potential given by \por.}
\label{fig:velcurv}
\end{figure}

To verify the intrinsic stability of our galaxy models, we run them in isolation at the resolution of $50 \pc$ and without star formation. \fig[s]{isomorpho_newton} and \ref{fig:isomorpho_mond} show the evolution of the face-on and edge-on morphology for the stellar and gaseous components. \fig{velcurv} displays the evolution of the velocity curves, computed using the net gravitational potential.

At the beginning (recall \sect{ic}), the gas component cools down into a thin disc. Once this stage has been reached, the ISM remains approximately the same. Because of this initial phase, we start all galaxy-galaxy models at least $100 \Myr$ before the first passage, allowing them to reach a relaxed stage before the interaction.

We note that the stellar component remains fairly stable over the duration of the integration ($600 \Myr$, i.e. several rotation periods), except for a mild enhancement of the density in the central $\approx 2 \kpc$, for both models, and the formation of a spiral instability in the Milgromian case. The gas follows the stellar potential well and also forms spiral arms. Although this introduces differences between the two models, their azimuthal averaged density profiles remain comparable.

\begin{figure}
\includegraphics{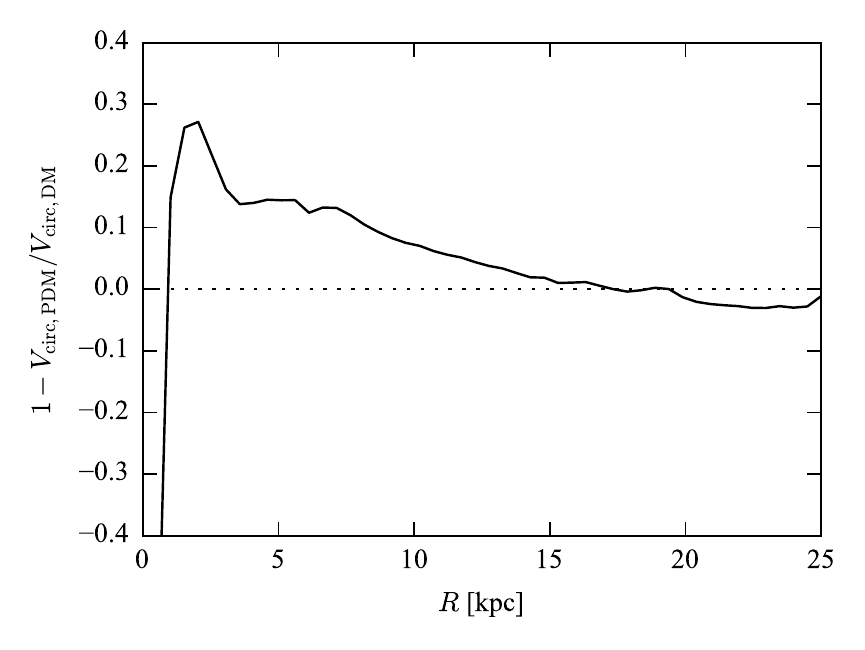}
\caption{Relative difference between the circular velocity from the phantom dark matter and that from the dark matter, from our initial conditions.}
\label{fig:velcurv_relative}
\end{figure}

\fig{velcurv_relative} shows the relative difference between the circular velocities of the isolated galaxy from the phantom dark matter in the MOND case and the dark matter in the Newtonian case. This figure represents the imperfections of our assumption consisting in replacing the DM with the MOND formalism. Except in the central kpc, the DM halo has a stronger gravitational influence on the baryons than the phantom DM, where differences can locally reach 30\%. In these regions, the stellar component dominates the mass budget of the galaxy (\fig{velcurv}). The differences however become much smaller at larger radii, where the non-baryonic component takes over. Such differences account for the enhanced formation of substructures in the Milgromian model noted above. The differences remaining relatively small, and because of the simplicity of the method allowing for a fair comparison of the two frameworks, we use these models for our simulations of interacting galaxies.

In short, for both models, the overall mass budget remains remarkably constant over several rotation periods, although the disc of the Milgromian model is prone to the formation of substructures after a few rotation periods only.

%%%%%%%%%%%%%%%%%%%%%%%%%%%%%%%%%%%%%%%%%%%%%%%%%%%%%%%%%%%%%%%%%%%%%%%%%%%%%%%%
\section{Extended halos}
\label{sec:extended}

\begin{figure}
\includegraphics{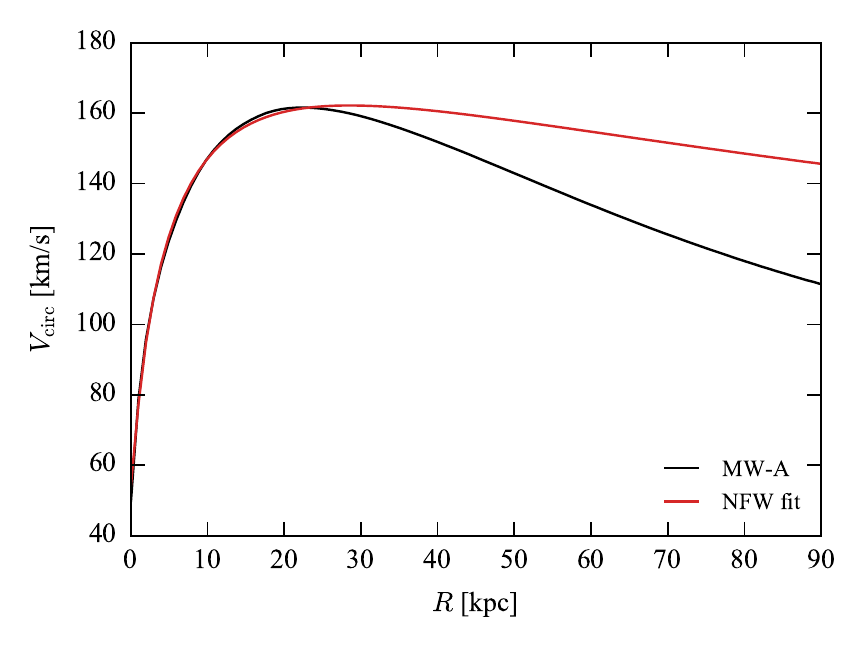}
\caption{Circular velocity of the MW-A model (truncated at $50 \kpc$) used in this work, and a fit within the inner $30 \kpc$ using a non-truncated NFW profile.}
\label{fig:nfwfit}
\end{figure}

Truncating DM halos represents a significant gain in memory and computational speed for simulations, and the model MW-A \citep{Kuijken1995} we use for our simulations takes advantage of such simplification. However in interacting galaxies, tidal tails extend to large distances and the gravitational potential they experience might be affected by the artificial truncation. In particular, \citet{Bournaud2003} shows that the formation of TDGs requires extended DM halo to support the fragmentation of the tails.

To ensure our results are not affected by the truncation artefact, we replace the lowered Evans halo with a more extended one, based on the NFW profile \citep{Navarro1997}. To do so, we fit the rotation curve of the naked DM halo of the MW-A model within the inner $30 \kpc$ with a NFW profile (\fig{nfwfit}). The best fitting NFW profile has a concentration of 13.7, a scale radius of $13.4 \kpc$ and a ``virial'' mass of $M_{200} \approx 6.65 \times 10^{11} \Msun$, leading to a ``virial'' radius of $r_{200} \approx 183 \kpc$. Using this set of parameters, we realise a complete galaxy model, keeping the baryonic components unchanged\footnote{For practical reasons, the new halo is truncated at $200 \kpc$.} and place it on the \newta orbit\footnote{The high mass of the new progenitor makes the interaction faster than in our fiducial model. To ensure our results on the effects of the extended halo are not affected by this change of the orbit, we also ran models with reduced velocities, and reach the same conclusions.}. Since the large-scale gas flows and small-scale cloud physics are not affected by the change of DM distribution in the outer parts of the halo, the SFR of this additional model is comparable within a few percent to that of the truncated models. The main differences occur at coalescence, likely because of the slight differences in the orbit induced by the higher total mass of the progenitors.

\begin{figure}
\includegraphics{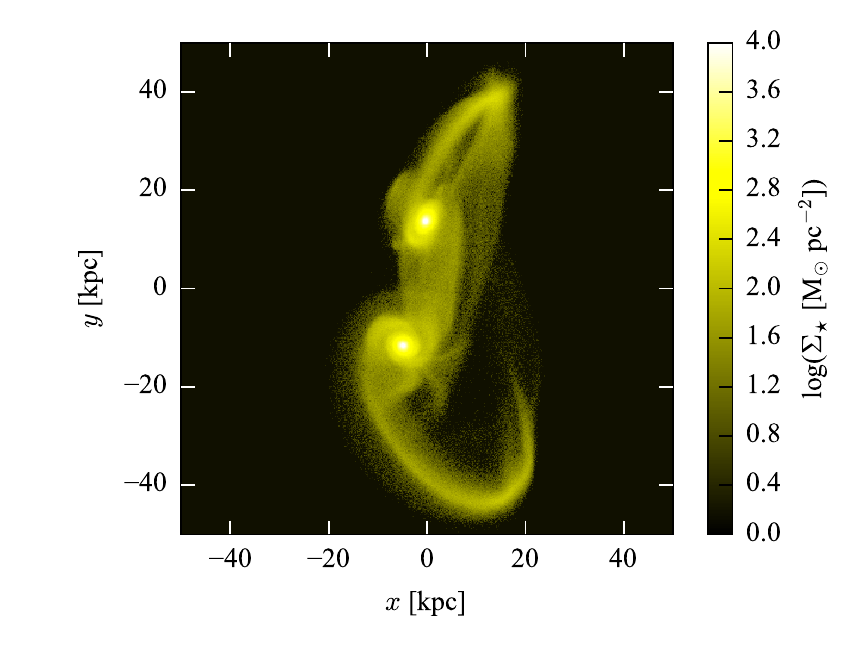}
\includegraphics{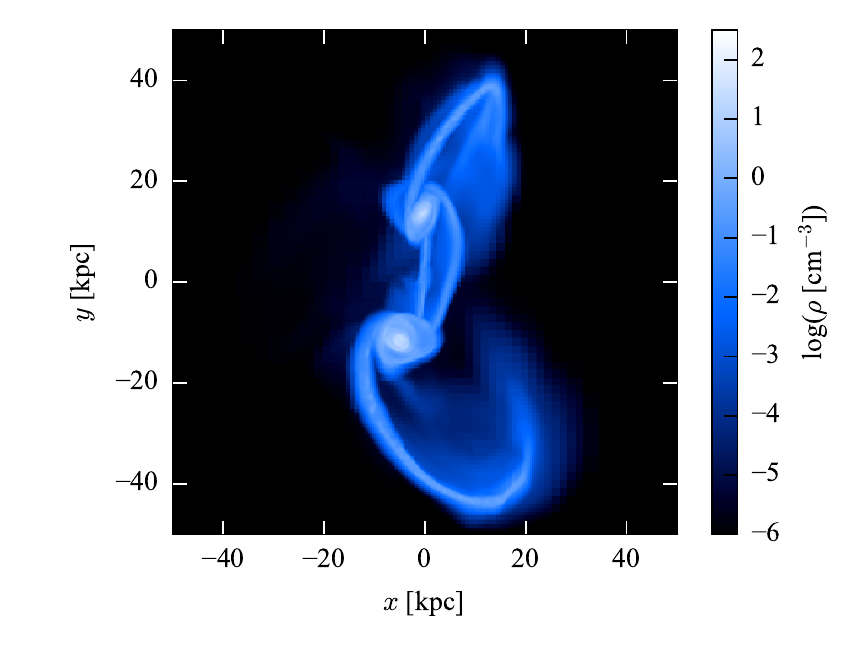}
\caption{Map of the stellar surface density (top) and the gas volume density (bottom) of the interacting model with extended halo. No substructure is visible in the tidal tails.}
\label{fig:morphoextended}
\end{figure}

\fig{morphoextended} shows that no substructure nor agglomeration is seen in the stellar or gaseous tidal tails, in a similar way as our more severely truncated fiducial model. We conclude that the possible formation of substructures in the tidal debris is not affected by our choice of truncation of the DM halo of our galaxy progenitors.

%%%%
\end{document}